\documentstyle[11pt]{article}
\newlength{\bredde}
\def\slash#1{\settowidth{\bredde}{$#1$}\ifmmode\,\raisebox{.15ex}{/}
\hspace*{-\bredde} #1\else$\,\raisebox{.15ex}{/}\hspace*{-\bredde} #1$\fi}
\newcommand{\beq}{\begin{equation}}
\newcommand{\eeq}{\end{equation}}
\def\beqn{\begin{eqnarray}}
\def\eeqn{\end{eqnarray}}
\def\sepand{\rule{14cm}{0pt}\and}
\def\gtwid{\raise.3ex\hbox{$>$\kern-.75em\lower1ex\hbox{$\sim$}}}
\def\ltwid{\raise.3ex\hbox{$<$\kern-.75em\lower1ex\hbox{$\sim$}}}
\begin{document}
%\begin{Ntitlepage}
\topmargin -1.4cm
\oddsidemargin -0.8cm
\evensidemargin -0.8cm
\title{\Large{
Symmetries and the Antibracket}}

\vspace{0.5cm}

\author{{\sc Jorge Alfaro} \\
Fac. de Fisica \\ Universidad Catolica de Chile\\
Casilla 306, Santiago 22, Chile \\
\sepand
{\sc Poul H. Damgaard}\thanks{On leave of absence from
the Niels Bohr Institute, Blegdamsvej 17, DK-2100 Copenhagen,
Denmark.} \\
Institute of Theoretical Physics\\Uppsala University,
Box 803\\S-751 08  Uppsala, Sweden}
\maketitle
\vfill
\begin{abstract} Requiring that a Lagrangian path integral leads to certain
identities (Ward identities in a broad sense) can be formulated in a
general BRST language, if necessary by the use of collective fields.
The condition of BRST symmetry can then be expressed with the help of
the antibracket, and suitable generalizations thereof. In particular, a
new Grassmann-odd bracket, which reduces to the conventional antibracket
in a special limit, naturally appears. We illustrate the formalism with
various examples.

\end{abstract}
\vfill
%\vspace{6.2cm}
\begin{flushleft}
UUITP-10/95 \\
hep-th/9505156
\end{flushleft}
\newpage
%\phantom{}
%\vfill
%\eject

%\setcounter{page}{1}

\section{Introduction}

In Lagrangian quantum field theory one normally starts with an action
depending on a set of classical fields, and then proceeds to quantize the
theory, often through the introduction of additional fields
(ghosts, ghosts-for-ghosts,
auxiliary fields, etc.). But the development can also proceed in the
inverse order. One may wish to have imposed, already at the quantum level,
a number of identities among Green functions, and is then seeking a
quantum action which, together with a specified functional measure, will
imply these identities. As examples of this kind, one can think of
chiral lagrangians (which lead to the correct chiral Ward identities, and
the associated current algebra), or quantized Yang-Mills theory (which ensures
the relations generically known as Ward identities).

Gauge theories of the most general kinds can all be quantized in
Lagrangian form using the
remarkable framework of Batalin and Vilkovisky \cite{Batalin}. What is
the underlying principle behind? It turns out that one can view it as the
BRST principle which imposes the most general identities of any quantum
field theory, the Schwinger-Dyson equations \cite{us}. Schwinger-Dyson
equations are normally not associated with any BRST symmetry, but such
a connection can be established by the help of collective fields
\cite{us0}. For a relation between the Batalin-Vilkovisky ``quantum
BRST operator" $\sigma$ and Schwinger-Dyson equations, see also ref.
\cite{Henneaux}. Imposing that the Lagrangian path integral be symmetric
with respect to the Schwinger-Dyson BRST operator \cite{us0} leads directly
to what is known as the Batalin-Vilkovisky (quantum) Master Equation
\cite{us}. This holds in the most general case, including that of
open gauge algebras.

A crucial ingredient in the Batalin-Vilkovisky scheme
is the so-called antibracket, an odd
Poisson-like bracket $(\cdot,\cdot)$ defined by
\beq
(F,G) = \frac{\delta^r F}{\delta\phi^A}\frac{\delta^l G}{\delta\phi^*_A}
-\frac{\delta^r F}{\delta\phi^*_A}\frac{\delta^l G}{\delta\phi^A}
\eeq
for a set of fields $\phi^A$ and ``antifields'' $\phi^*_A$ that are canonically
conjugate within this bracket,
\beq
(\phi^A,\phi^*_B) = \delta^A_B;(\phi^A,\phi^B) = (\phi^*_A,
\phi^*_B) = 0 .
\eeq
For flat functional measures, the Master Equation for the quantum action
can be expressed entirely in terms of the antibracket and a quadratic
nilpotent operator
\beq
\Delta \equiv (-1)^{\epsilon_A+1}\frac{\delta^r}{\delta\phi^A}
\frac{\delta^r}{\delta\phi^*_A} .
\eeq
Here $\epsilon_A$ is the Grassmann parity of the fields $\phi^A$. The
antifields $\phi^*_A$ have Grassmann parities $\epsilon_A+1$. (If one does
not consider flat measures for the fields $\phi^A$, the appropriate object
is a covariant generalization of $\Delta$ discussed in ref.
\cite{Schwarz,us1}; see also ref. \cite{HT}.) Because of this difference
in Grassmann parities, the
antibracket does not share many properties with the usual Poisson bracket
(and its fermionic analogue). In the anitbracket formalism the ``momenta''
are quite different from ``coordinates'', and this extends also to, $e.g.$,
ghost number assignments.

It is very simple to derive a Master Equation the quantum action $S_{ext}$
must satisfy for {\em any} theory with flat measure. One starts with
the Schwinger-Dyson BRST symmetry \cite{us0}
\beq
\delta \phi^A = c^A~,~~~
\delta c^A = 0~,~~~
\delta \phi^*_A = - \frac{\delta^l S}{\delta\phi^A} ~.
\eeq
involving the fields $\phi^A$ and a conventional pair of ghosts $c^A$ and
antighosts (the antifields of the Batalin-Vilkovisky formalism) $\phi^*_A$.
Invariance of the path integral with respect to this symmetry means that
the BRST variation of the action is cancelled by the Jacobian from the
measure. This immediately implies
\beq
\delta S_{ext} = \frac{1}{2}(S_{ext},S_{ext}) + \frac{\delta^r S_{ext}}
{\delta\phi^A}c^A = i\hbar \Delta S_{ext}.
\eeq
Or, in terms of $\psi \equiv \exp[(i/\hbar)S_{ext}]$,
\beq
i\hbar\Delta\psi = c^A\frac{\delta^r}{\delta\phi^A}\psi .
\eeq

The substitution $S_{ext} = S^{BV} - \phi^*_Ac^A$ leads to the quantum
Master Equation of the Batalin-Vilkovisky formalism:
\beq
\frac{1}{2}(S^{BV},S^{BV}) = i\hbar \Delta S^{BV} .
\eeq

Some natural questions arise in this connection. If the most general
requirement
for any quantum field theory ---  that correct Schwinger-Dyson equations are
reproduced by functional averaging of Green functions ---  leads to the most
general BRST quantization prescription, what happens if we impose more
stringent conditions? Can it be phrased in BRST language, and, if so, will
the antibracket still play an important r\^{o}le? Can we impose different
Master Equations to be satisfied by new classes of actions in the
same manner as the Batalin-Vilkovisky Master Equation follows from demanding
Schwinger-Dyson BRST symmetry?

These questions will be answered as we proceed. One of the main
lessons is that the conventional antibracket formalism, being rooted
in the a priori assumption of a one-to-one matching between fields and
antifields, is a very special case. Figuratively speaking, by having
one antifield for each field, one has completely fixed the quantum
dynamics, modulo boundary conditions, of the theory. Technically, this
manifests itself in the fact that all Schwinger-Dyson equations are
reproduced by the BRST operator. If the number of
fields does not match that of the antifields
(a misleading name in this context; they are then just a particular set
of ordinary antighosts of a certain BRST symmetry one wishes to
impose), the quantum
dynamics is not uniquely specified by the associated Master Equation.
In such a case there is no natural symplectic structure in the
formalism, and there is not an equal number of ``coordinates'' and
``momenta'' with which to define a canonical bracket. We emphasize that these
generalized settings nevertheless are as valid as those based on the
full set of Schwinger-Dyson equations. They still define classes of
theories with certain specified BRST symmetries.

It is thus important to realize that the canonical structure and the
antibracket itself is not required to define classes of quantum
actions. The antibracket appears in the ultimate case, where the
quantum theory is completely specified by the associated BRST
symmetry. In all other cases some of the ``momenta'' $\phi^*$ are
missing from the formalism, which is only defined on a truncated phase space.
The nature of the pertinent smaller
space of fields and antighosts $\phi^*$ can
be derived from the same collective field formalism that is used to
derive the Batalin-Vilkovisky scheme, and cannot be obtained by any
naive truncations of the set of antighosts. The associated generalizations
thus cannot be inferred without this knowledge on how to
derive the Batalin-Vilkovisky formalism from a more fundamental
principle \cite{us}. Whether the existence of suitable generalizations
of the Batalin-Vilkovisky Lagrangian formalism can be of use in new
contexts remains to be explored (and in particular, see the comments below).
We shall here restrict ourselves
to deriving the main principles, and to show how known results can be
understood in this new light.

What could be a motivation for trying to define classes of field theories on
the basis of an underlying Master Equation for the quantum action? In the
conventional antibracket formalism the answer is that it provides a
systematic approach to BRST gauge fixing. All of the physical dynamics
resides in the classical action and in the functional measure of the
classical fields. In the Batalin-Vilkovisky formalism, one finds an
extended action which leads to the same Schwinger-Dyson equations that
one formally derives in the original theory, before gauge fixing. The
advantage of this extended action is that it immediately can be brought in
a form where the gauge has been fixed correctly, and where the path
integral thus takes on a less formal significance. Similarly, even if we can
find non-trivial solutions to Master Equations that lead to a more
restricted set of Ward Identities, such generalizations mainly have
applications at the level of gauge fixing. The classical dynamics will
always be used as a boundary condition, and in this sense the Master
Equations do not define for us broader classes of classical actions that
will lead to the same quantum dynamics. However, and not surprisingly,
the Master Equations will tell us which are the local internal symmetries
of the classical action that are compatible with the required set of Ward
Identities. Only in this restricted meaning does the new Master Equations
stipulate conditions to be imposed on the classical actions.

The paper is organized as follows. In the next section we consider the
derivation of a covariant formulation of the usual Batalin-Vilkovisky
formalism, using the same procedure as in ref. \cite{us}. (The only new
ingredient is the presence of a non-trivial scalar measure density
$\rho(\phi)$ in the functional integral). We show how to derive a
corresponding Schwinger-Dyson BRST symmetry, and focus on the non-Abelian
version of this formulation.  Section 3 is devoted to most general
setting possible: The case
of field transformations that leave neither measure nor action invariant,
and in section 4 we discuss a new Grassmann-odd bracket structure that
naturally emerges in this context. Section 5 contains our conclusions.
In two appendices we illustrate the BRST
technique by deriving the analogue of a Batalin-Vilkovisky formalism
for theories defined in terms of group variables of a given Lie
group.

\vspace{1cm}

\section{Theories with non-trivial measures}

As a first illustration of the generality of the BRST approach to
implementing chosen identities in the path integral, we shall here
consider the result of imposing symmetries of a
given functional measure in the form of BRST invariance. A brief
collection of the main results of this section have been reported in ref.
\cite{us1}, but no details of the derivation were given there.

Consider a classical action $S_{cl}$ of classical fields $\phi^A$.
Suppose we wish to
quantize this theory in the path integral framework by integrating over
a certain functional measure $[d\phi]\rho(\phi)$. Naively,
the result would be a partition function of the form
\beq
{\cal{Z}} = \int [d\phi]\rho(\phi)e^{\frac{i}{\hbar} S_{cl}[\phi]} ~,
\eeq
but we know that this prescription often, and in particular if we have
internal gauge symmetries, is inadequate. Obtaining correct Schwinger-Dyson
equations can again be used as the principle to enlarge the path integral
in a proper way. In the case of a flat functional measure, the Schwinger-Dyson
equations were obtained by exploiting the symmetry of the measure under
local translations, $\phi^A(x) \to \phi^A(x) + \varepsilon^A(x)$. Despite
the extra factor of $\rho(\phi)$ in the measure (8), we can still make
use of the invariance of just $[d\phi]$ in this case. This leads, for an
arbitrary functional $F$, to one
particular form of the Schwinger-Dyson equations:
\beq
\left\langle (-1)^{\epsilon_M}\Gamma^M_{AM}(\phi)F(\phi) + \left(\frac{i}
{\hbar}\right)\frac{\delta^l S}{\delta\phi^A}F(\phi) + \frac{\delta^l F}
{\delta\phi^A}\right\rangle = 0 ,
\eeq
with
\beq
(-1)^{\epsilon_M}\Gamma^M_{AM} \equiv \rho^{-1}\frac{\delta^l\rho}
{\delta\phi^A} .
\eeq

A more natural definition of Schwinger-Dyson equations may appear to be
based on the full measure $[d\phi]\rho(\phi)$. Whereas the interesting
invariance of the part $[d\phi]$ alone is a local (flat) translation, the
analogous transformation in the case of $[d\phi]\rho(\phi)$ will in
general take place on a non-trivial space. Subtleties can certainly arise
when this space is multiply-connected, has boundaries, etc., but we shall
here restrict ourselves to the case where the group of transformations
which leave this full measure invariant is continuous. We are then in
general dealing with a super Lie group, whose generators shall be given
below.

Denote the set of transformations that leave $[d\phi]\rho(\phi)$ invariant
by $g$:
\beq
\phi^A(x) = g^A(\phi'(x),a(x)),
\eeq
where $a^i(x)$ are local fields parametrizing the transformation. They are
the analogues of the local translations in the case of a flat measure. The
transformation $g$ must obviously be connected to the identity, which we
take to occur at $a^i(x)=0$:
\beq
g^A(\phi'(x),0) = \phi'^A(x).
\eeq

It can happen that also the action $S$ is invariant under the set of
transformations
(11). This is a very special case for which, in the absence of spontaneous
symmetry breaking, classical equations of motion
suffice to define the full quantum theory. Certain topological field
theories fall into this class. But
in general the transformation (11) will not be a symmetry of the action.

Making the transformation (11) close to the identity leads to the following
relations, for an arbitrary functional $G$:
\beq
\left\langle v^B_i(\phi)\left[\frac{\delta^l G}{\delta\phi^B} + \left(\frac{i}
{\hbar}\right)\frac{\delta^l S}{\delta\phi^B}G(\phi)\right]\right\rangle = 0,
\eeq
where
\beq
v^A_i(\phi) \equiv \left.\frac{\delta^l g^A}{\delta a^i}\right|_{a=0} .
\eeq

These identities appear to differ from the Schwinger-Dyson equations (9).
Consider, however, the case where there are locally just as many parameters
$a^i(x)$ of
the group as there are local translations. In particular, a one-to-one
relation
can be established with the Grassmann parities of the parameters $a^A$
and the original fields $\phi^A$, $i.e.$, $\epsilon(a^A) =
\epsilon(\phi^A)$. Now start with the original
Schwinger-Dyson equations (9). Since they hold for an arbitrary $F$, they
hold in particular for the choice $F^A_B(\phi) = (-1)^{\epsilon_B
(\epsilon_A+1)}v^B_A(\phi)G(\phi)$. Substituting this into (9), and
summing over $B$, one finds:
\beq
\left\langle v^B_A(\phi)\left[\frac{\delta^l G}{\delta\phi^B} + \left(\frac{i}
{\hbar}\right)\frac{\delta^l S}{\delta\phi^B}\right] + (-1)^{\epsilon_M}
v^B_A(\phi)\Gamma^M_{BM}(\phi)G(\phi)
+ \frac{\delta^r v^B_A}{\delta\phi^B}
G(\phi)\right\rangle = 0 .
\eeq
So if one can show that
\beq
\frac{\delta^r v^B_A}{\delta\phi^B} + (-1)^{\epsilon_M}v^B_A(\phi)
\Gamma^M_{BM}(\phi) = 0,
\label{16}
\eeq
then the two sets of equations will be equivalent. How could an identity
such as (16) arise? Our only information is that the measure $[d\phi]
\rho(\phi)$ is invariant under the transformation (11). Making the
substitution (11), and expanding around the identity transformation, i.e.,
$\phi^A = \phi'^A + a^Bv^A_B(\phi) + \ldots$, we get
\beq
(-1)^{\epsilon_A}\rho(\phi)\frac{\delta^r v^A_B}{\delta\phi^A}
+ v^A_B(\phi)\frac{\delta^l \rho}{\delta\phi^A} = 0 .
\eeq
If $v$ is invertible, this can be written
\beq
\frac{\delta^l\rho}{\delta\phi^C} + (-1)^{\epsilon_A}\rho(\phi)
\left(v^{-1}\right)^B_C\frac{\delta^r v^A_B}{\delta\phi^A} = 0 .
\eeq
But this is precisely the relation (16), provided we identify
\beq
(-1)^{\epsilon_A}\Gamma^A_{CA} = (-1)^{\epsilon_A+1}\left(v^{-1}\right)^B_C
\frac{\delta^r v^A_B}{\delta
\phi^A} ,
\eeq
which in turn is just the condition that (\ref{16}) is satisfied. So under the
above
conditions the two sets of Schwinger-Dyson equations (9) and (13) are
equivalent.

\subsection{BRST formulations}

Next, we impose the Schwinger-Dyson equations as Ward identities of an
unbroken BRST symmetry. For theories with flat measures, this has been
explained in ref. \cite{us0}. When the measure is of the form $[d\phi]
\rho(\phi)$, there are several routes to a proper BRST description. The
simplest is to just blindly exponentiate the measure density $\rho(\phi)$,
and treat this as a one-loop quantum correction to the action. We can
then obviously employ the formalism of the flat measure, provided we
substitute
\beq
S[\phi] \to S[\phi] - i\hbar \int dx \ln[\rho(\phi)]
\eeq
into eq. (4). This gives
\begin{eqnarray}
\delta\phi^A &=& c^A \cr
\delta c^A &=& 0 \cr
\delta\phi^*_A &=& - \frac{\delta^l S}{\delta\phi^A} + i\hbar
(-1)^{\epsilon_M}\Gamma^M_{AM} ,
\end{eqnarray}
and the Ward identities $0 = \langle\delta(\phi^*_A F[\phi])\rangle$ are
then the Schwinger-Dyson equations (9). In this way one can evidently
always proceed with the quantization procedure, just as in ref.
\cite{us}. But done in this way one has clearly lost all of the geometric
interpretation associated with quantizing a theory on a non-trivial
field space.

An alternative BRST formulation follows the second way of deriving
Schwinger-Dyson equations, as described above. One here exploits the
symmetry of the full measure $[d\phi]\rho(\phi)$, and promotes this symmetry
into a local gauge invariance (of both measure and action). The BRST Ward
identities will then be Schwinger-Dyson equations of the form (13)
\cite{us2}. Some of the manipulations to be considered below are
clearly going to be formal in the sense that they will ignore possible
modifications due to regularization. At the two-loop level certain
field transformations may also require some modifications of the
relevant BRST algebra \cite{us3}.

In the Abelian formulation, integrate with flat measures
over the collective fields $a^A(x)$, the auxiliary fields $B_A(x)$, and
the ghost--antighost pair $c^A(x),\phi^*_A(x)$. The functional integral
is then invariant under the nilpotent BRST symmetry
\begin{eqnarray}
\delta\phi'^A &=& - \left(M^{-1}\right)^A_B
\frac{\delta^r g^B}{\delta a^C}c^C \cr
\delta a^A &=& c^A \cr
\delta c^A &=& 0 \cr
\delta\phi^*_A &=& B_A \cr
\delta B_A &=& 0
\label{22}
\end{eqnarray}
where
\beq
M^A_B(\phi',a) \equiv \frac{\delta^r g^A}{\delta\phi'^B} ~.
\label{23}
\eeq

Gauge-fixing the collective field $a$ to zero is achieved by adding a term
$-\delta[\phi^*_Aa^A] = (-1)^{\epsilon_A+1}B_Aa^A - \phi^*_Ac^A$ to the
action $S$. After integrating out $a$ and $B$, one is left with the following
transformations:
\begin{eqnarray}
\delta\phi^A &=& - u^A_B(\phi)c^B \cr
\delta c^A &=& 0 \cr
\delta\phi^*_A &=& (-1)^{\epsilon_A+\epsilon_B}
\frac{\delta^l S}{\delta\phi^B}u^B_A(\phi) ,
\end{eqnarray}
where we have used eq. (\ref{23}), and in which
\beq
u^A_B(\phi) \equiv \left.\frac{\delta^r g^A}{\delta a^B}\right|_{a=0} .
\label{25}
\eeq

Let us for the moment assume that $u^A_B$ is invertible, and let us
then perform a simple redefinition:
\beq
C^A = - u^A_B(\phi)c^B, \Phi^*_A = -\phi^*_B\left(u^{-1}
\right)^B_A .
\eeq
Barring anomalies associated with the ghost--antighost measure $[dc]
[d\phi^*]$, this transformation has unit Jacobian.
In terms of these new variables, the BRST transformations (24) become:
\begin{eqnarray}
\delta\phi^A &=& C^A \cr
\delta C^A &=& (-1)^{\epsilon_A\epsilon_D}\Gamma^A_{DC}C^CC^D \cr
\delta\Phi^*_A &=& (-1)^{\epsilon_D+1}\Gamma^D_{AC}C^C\Phi^*_D
- \frac{\delta^l S}{\delta\phi^A} .
\label{27}
\end{eqnarray}

We have here introduced a natural connection associated with the measure
$[d\phi]\rho(\phi)$, namely
\beq
\Gamma^D_{AC} \equiv (-1)^{\epsilon_A(\epsilon_D+1)}u^D_B(\phi)
\frac{\delta^r\left(u^{-1}\right)^B_A}{\delta\phi^C} .
\eeq

One can readily check that this definition is compatible with eq. (19). Now,
since the original functional measure was invariant under the
transformation (11), the same should be the case for the formulation (27).
Indeed, one finds that the new measure $[d\phi][dC][d\Phi^*]\rho(\phi)$
is invariant under (27) precisely if
\beq
\frac{\delta^r\rho}{\delta\phi^D} - (-1)^{\epsilon_A+\epsilon_D}\rho
\Gamma^A_{DA} = 0 ,
\eeq
which is just the condition that the original field measure $[d\phi]
\rho(\phi)$ is invariant under (11).

What about nilpotency of the BRST transformation (27)? By construction, our
BRST transformations are always nilpotent when all fields of the formalism
are included. When we start to integrate out part of these fields, we will
in general lose nilpotency. In the present case one can easily check that
nilpotency of $\delta$ in general is lost already at the level where it
acts only on the original fields: $\delta^2\phi^A \neq 0$\footnote{Nilpotency
is preserved in the case where the group of transformations is Abelian, but
this is of course a very special case.}. Although there is nothing wrong
with such a formalism, it makes it very difficult to use it as a basis for
a quantization programme. We will therefore instead focus on an alternative
formulation, described below.

\subsection{The non-Abelian formulation}

Since the set of transformations that leave the measure invariant
most often will form a non-Abelian (super) Lie group, a more natural
formulation of the associated Schwinger-Dyson BRST symmetry arises
from the corresponding non-Abelian local gauge symmetry introduced by
the help of collective fields. We shall here give a few details
related to this formulation.

First a few definitions. Since the transformations (11) form a
(super) group, two consecutive transformations parametrized by local
fields $a^A(x)$ and $b^A(x)$ must be expressible as a single
transformation of some new parameters, let us denote them by
$$
\psi^A(b,a) .
$$
In detail, let
\beq
\phi^A = g^A(\phi',a) {\mbox{\rm and}} \phi'^A =
g^A(\phi'',b) ,
\eeq
then
\beq
\phi^A = g^A(g(\phi'',b),a) = g^A(\phi'',\psi(b,a)) .
\eeq

Next, differentiate this equation on both sides with respect to $b^A$,
set $b^A=0$, and use the boundary conditions
\beq
\psi^A(0,a) = a^A  {\mbox{\rm and}}  \phi'^A = g^A(\phi'',0) =
\phi''^A
\eeq
to get
\beq
\frac{\delta^r g^A(\phi',a)}{\delta\phi'^B}u^B_C(\phi') = \frac{\delta^r
g^A(\phi',a)}{\delta a^B}\nu^B_C(a) .
\eeq
Here we have introduced
\beq
\nu^A_B(a) \equiv \left.\frac{\delta^r\psi^A(b,a)}{\delta
b^B}\right|_{b=0} ,
\eeq
which has an inverse, $\lambda$, defined by
\beq
\lambda^A_B\nu^B_C = \delta^A_C .
\eeq

Consider now
\beq
\delta\phi^A = \frac{\delta^r \phi^A(\phi',a)}{\delta\phi'^B}
\delta\phi'^B + \frac{\delta^r \phi^A(\phi',a)}{\delta a^B}\delta a^B.
\label{36}
\eeq
It follows from (\ref{36}) that if choose
\begin{eqnarray}
\delta\phi'^A &=& \left.\frac{\delta^r \phi^A(\phi',a)}{\delta
a^B}\right|_{a=0}\varepsilon^B = u^A_B(\phi)\varepsilon^B \cr
\delta a^A &=& -\nu^A_B(a)\varepsilon^B
\label{37}
\end{eqnarray}
then the original field $\phi^A$ is left invariant. So this is a local
gauge symmetry of the transformed action. This can also be derived in a
more conventional manner by starting from the Hamiltonian formulation
of the collective field formalism, and then multiplying the symmetry
generators by an appropriate combination of fields \cite{Hosoya}.

So far we have only verified that the gauge transformations (37)
generate a symmetry of the action. Will it be a genuine
quantum-mechanical symmetry in the path integral? This clearly depends
on the choice of measure for the collective fields $a^A$. (The measure
$[d\phi]\rho(\phi)$ is, by construction, invariant under the
$\phi$-transformation above). As discussed in ref. \cite{us2}, the
functional measure for $a^A$ will be invariant if we choose it to be
an either left or right invariant (Haar) measure. In the following
we shall consider the left invariant Haar measure.

The gauge symmetry
(37) above is the non-Abelian analogue of the Abelian symmetry (27).
As in all such formulations, there may be situations where there are
obstructions to such a shift between Abelian and non-Abelian
formulations of the same underlying gauge symmetry. What is important
for our purposes is that the non-Abelian gauge symmetry (37) incorporates
the conventional Schwinger-Dyson equations as BRST Ward identities. To
see this, we first have to introduce the analogous BRST symmetry, and
make use of some fundamentals of (super) group theory. Consider first
the BRST transformations corresponding to the gauge symmetry (37):
\begin{eqnarray}
\delta\phi'^A &=& u^A_B(\phi')c^B \cr
\delta a^A &=& -\nu^A_B(a)c^B .
\end{eqnarray}
Imposing nilpotency of this BRST operator $\delta$ fixes the
transformation law for the ghosts $c^A$. After imposing
$\delta^2\phi'^A = 0$, one finds:
\beq
\delta c^A = (-1)^{\epsilon_B}\left(u^{-1}\right)^E_A\frac{\delta^r
u^A_B}{\delta \phi'^C} u^C_D c^D c^B .
\label{39}
\eeq

This transformation of the ghosts is, as usual for non-Abelian gauge
symmetries, directly related to the structure coefficients $c^A_{BC}$
for the (super) group. One has \cite{Hamermesh}:
\beq
\frac{\delta^r u^A_B}{\delta\phi'^C}u^C_D -
(-1)^{\epsilon_B\epsilon_D}\frac{\delta^r u^A_D}{\delta \phi'^C}u^C_B
= - u^A_C c^C_{BD} .
\label{40}
\eeq
The structure ``coefficients'' are supernumbers with the property
\beq
c^C_{BD} = - (-1)^{\epsilon_B\epsilon_D}c^C_{DB} .
\eeq

With the help of eq. (40), one can rewrite the BRST transformation for
the ghosts as
\beq
\delta c^A = - \frac{1}{2}(-1)^{\epsilon_B}c^E_{BD}c^Dc^B .
\eeq
Note that
\beq
U^E_{BD} \equiv (-1)^{\epsilon_B}c^E_{BD}
\eeq
has the following symmetry property:
\beq
U^E_{BD} = (-1)^{(\epsilon_B+1)(\epsilon_D+1)} U^E_{DB} ,
\eeq
which precisely is what is required in order that the right hand side
of eq. (40), in general, is non-vanishing for arbitrary Grassmann parity
assignments of the ghosts.

It still remains to be checked whether the ghost transformation law (39)
is compatible with nilpotency of the BRST charge. One finds that indeed
$\delta^2 c^A\!=\!0$ as a consequence of the generalized Jacobi identity
(see the second reference in \cite{Hamermesh}):
\begin{eqnarray}
(-1)^{\epsilon_B\epsilon_E}c^A_{BF}c^F_{CE} + (-1)^{\epsilon_C
\epsilon_E}c^A_{EF}c^F_{BC} + (-1)^{\epsilon_B\epsilon_C}c^A_{CF}c^F_{EB}
= 0 .
\label{45}
\end{eqnarray}

The transformation (39) came from requiring $\delta^2\phi'^A = 0$. For
consistency we ought to obtain the same condition from imposing
nilpotency of $\delta$ when acting on the collective fields $a^A$.
This is indeed the case, but it interestingly turns out to involve non-trivial
identities from (super) group theory. In fact, these identities can be
{\em derived} from demanding a consistent BRST formulation. Demanding
$\delta^2 a^A=0$ leads to
\beq
\delta c^E = (-1)^{\epsilon_B+1}\lambda^E_A\frac{\delta^r
\nu^A_B}{\delta a^C}\nu^C_D(a)c^Dc^B .
\eeq

Can (46) be consistent with (39)? Imposing the (super) integrability
condition
\beq
\frac{\delta^r \delta^r \phi^A}{\delta a^C \delta a^B} =
(-1)^{\epsilon_B\epsilon_C} \frac{\delta^r \delta^r \phi^A}{\delta a^B
\delta a^C} ,
\eeq
and using the composition property
\beq
\frac{\delta^r \phi^A}{\delta a^B} = - u^A_C(\phi)\lambda^C_B(a) ,
\eeq
we obtain
\beq
u^A_D\left[\frac{\delta^r\lambda^D_B}{\delta a^C} -
(-1)^{\epsilon_B\epsilon_C} \frac{\delta^r \lambda^D_C}{\delta
a^B}\right]
-(-1)^{\epsilon_C(\epsilon_B+\epsilon_D)}\left[\frac{\delta^r
u^A_D}{\delta \phi^E} u^E_F -
(-1)^{\epsilon_D\epsilon_F}\frac{\delta^r u^A_F}{\delta \phi^E} u^E_D
\right] \lambda^F_C\lambda^D_B  = 0 .
\eeq

Combining this with eq. (40) leads to the analogue of that equation,
now expressed in terms of variables derived from the parameters of the
group rather than the group coordinates themselves:
\beq
\frac{\delta^r \lambda^G_C}{\delta a^B} - (-1)^{\epsilon_B\epsilon_C}
\frac{\delta^r \lambda^G_B}{\delta a^C} = (-1)^{\epsilon_C\epsilon_D}
c^G_{DF} \lambda^F_C\lambda^D_B .
\eeq
This relation, eq. (50), is precisely what is needed to show that the
ghost transformation (46) is equivalent to eq. (39).

To summarize this part, we have succeeded in setting up the consistent
BRST multiplet associated with the collective-field gauge symmetry
(37). When supplemented with a conventional antighost $\phi^*_A$
associated with the ghost $c^A$, and an auxiliary field $B_A$, we can
summarize these BRST transformations below:
\begin{eqnarray}
\delta \phi'^A &=& u^A_B(\phi')c^B \cr
\delta a^A &=& - \nu^A_B(a)c^B \cr
\delta c^A &=& - \frac{1}{2} (-1)^{\epsilon_B}c^A_{BC}c^Cc^B \cr
\delta \phi^*_A &=& B_A \cr
\delta B_A &=& 0 .
\label{51}
\end{eqnarray}

As before, the trick is now to integrate out the collective fields
$a^A$ by choosing an appropriate gauge. It turns out to be convenient
to introduce an object
\beq
\bar{\Gamma}^A_{BC} \equiv \left.\frac{\delta^r\nu^A_B}{\delta
a^C}\right|_{a=0} .
\eeq
Let us first consider some of its properties. From eq. (50) it follows
that
\beq
\lambda^G_B\frac{\delta^r\nu^B_K}{\delta a^C}\nu^C_L -
(-1)^{\epsilon_K\epsilon_L}\lambda^G_B\frac{\delta^r\nu^B_L}{\delta a^C}
\nu^C_K = c^G_{KL} .
\eeq
When evaluated at $a^A\!=\!0$, this relation implies
\beq
\bar{\Gamma}^G_{KL} - (-1)^{\epsilon_K\epsilon_L}\bar{\Gamma}^G_{LK}
= c^G_{KL} .
\eeq

The object $\bar{\Gamma}^A_{BC}$ plays a role in the non-Abelian
Schwinger-Dyson BRST transformations, once the collective fields $a^A$
(and auxiliary fields $b_A$) have been integrated out. We first
gauge-fix all collective fields $a^A$ to zero by adding to the action
a BRST gauge-fixing term of the form
\beq
-\delta\{\phi^*_Aa^A\} = (-1)^{\epsilon_A+1}B_Aa^A +
\phi^*_A\nu^A_B(a)c^B .
\eeq
Note that the ghost--antighost term is non-trivial in this
formulation.

Next, let us integrate out the fields $a^A$ and $b_A$. As discussed
above, we are required to take a (super) Haar measure for $a$. We can
write it as $[da]_E {\mbox{\rm sdet}}[\lambda(a)]$, where the subscript $E$
denotes the flat (euclidean) measure, and ``sdet'' denotes the
superdeterminant. This means that the analogue of the equation of
motion for $a^A$ will contain a quantum contribution as well:
\beq
\frac{\delta^r S}{\delta a^M} - i\hbar\frac{\delta^r}{\delta a^M}
\ln\left[{\mbox{\rm sdet}}(\lambda(a))\right] + (-1)^{\epsilon_M+1}B_M
+ (-1)^{\epsilon_M(\epsilon_B+1)}\phi^*_A\frac{\delta^r
\nu^A_B}{\delta a^M} c^B = 0 .
\eeq
Now,
\begin{eqnarray}
\frac{\delta^r}{\delta a^M}\ln\left[{\mbox{\rm
sdet}}(\lambda(a))\right] = -\frac{\delta^r}{\delta
a^M}\ln\left[{\mbox{\rm sdet}}(\nu(a))\right] &=& -{\mbox{\rm sTr}}
\left[\nu^{-1}(a)\frac{\delta^r \nu(a)}{\delta a^M}\right] \cr
&=& \sum_A (-1)^{\epsilon_A+1}\lambda^A_B(a)
\frac{\delta^r\nu^A_B(a)}{\delta a^M} ,
\end{eqnarray}
which, when evaluated at $a\!=\!0$, gives
$$
(-1)^{\epsilon_A+1}\bar{\Gamma}^A_{AM} .
$$

So the ``quantum mechanical equation of motion'' for $B$,
evaluated at $a\!=\!0$, becomes
\beq
B_M = (-1)^{\epsilon_M}\frac{\delta^r S}{\delta\phi^A} u^A_M(\phi) +
i\hbar (-1)^{\epsilon_A+\epsilon_M}\bar{\Gamma}^A_{AM} +
(-1)^{\epsilon_M\epsilon_B}\phi^*_A\bar{\Gamma}^A_{BM}c^B .
\eeq
Using the boundary condition $g^A(\phi',a\!\!=\!\!0)\!=\!\phi'^A$, this
means that the BRST transformations (51) turn into
\begin{eqnarray}
\delta\phi^A &=& u^A_B(\phi)c^B \cr
\delta c^A &=& -\frac{1}{2}(-1)^{\epsilon_B}c^A_{BC}c^Cc^B \cr
\delta\phi^*_A &=& (-1)^{\epsilon_A}\frac{\delta^r S}{\delta\phi^B}
u^B_A(\phi) + i\hbar (-1)^{\epsilon_A+\epsilon_B}\bar{\Gamma}^B_{BA} +
(-1)^{\epsilon_A\epsilon_B}\phi^*_M\bar{\Gamma}^M_{BA}c^B .
\end{eqnarray}

The crucial test of the above BRST symmetry is to see if the associated Ward
identities are correct Schwinger-Dyson equations. We check it by
evaluating $0\!=\!\langle\delta[\phi^*_AF(\phi)]\rangle$ at the intermediate
stage where the ghost--antighost pair $c^A,\phi^*_A$ has been
integrated out. Note that the ghost expectation values in this case
have to be evaluated with respect to the extended action
\beq
S_{ext} = S[\phi] + \phi^*_Ac^A ,
\eeq
with a sign difference as compared with the Abelian formulation
\cite{us}.\footnote{And in particular, we now have the ghost expectation value
$\langle c^A\phi^*_B\rangle\!=\!+i\hbar\delta^A_B$.} After a number of
cancellations, one verifies that the above Ward identities turn into
\beq
\left\langle
(-1)^{\epsilon_A(\epsilon_B+1)}u^B_A(\phi)\left[\frac{\delta^l
F}{\delta \phi^B} + \left(\frac{i}{\hbar}\right)\frac{\delta^l
S}{\delta \phi^B}F(\phi)\right]\right\rangle = 0 ,
\eeq
which coincide with (13).

We can write the above BRST symmetry in a more compact manner by
making the redefinitions
\beq
C^A \equiv u^A_B(\phi)c^B,\Phi^*_A \equiv
\phi^*_B\left(u^{-1}\right)^B_A ,
\label{62}
\eeq
which, as in the Abelian formulation, is a transformation of unit
Jacobian, provided there are no anomalies associated with such a ghost
transformation.

In terms of these new variables,
\beq
\delta \phi^A = C^A ,
\eeq
so nilpotency requires $\delta C^A\!=\!0$. One can check that this
indeed automatically is satisfied when the ghost redefinition is as
given above. Finally, we can derive the transformation law for the
redefined antighost $\Phi^*_A$. Defining
\beq
\tilde{\Gamma}^M_{AK} \equiv
(-1)^{\epsilon_A(\epsilon_M+1)}\left\{u^M_B(\phi)
\frac{\delta^r\left(u^{-1}\right)^B_A}{\delta \phi^K} +
(-1)^{\epsilon_A\epsilon_C}
u^M_S(\phi)\bar{\Gamma}^S_{CB}\left(u^{-1}\right)^B_A
\left(u^{-1}\right)^C_K\right\} ,
\eeq
we can summarize the resulting BRST transformation for all remaining
fields:
\begin{eqnarray}
\delta\phi^A &=& C^A \cr
\delta C^A &=& 0 \cr
\delta\Phi^*_A &=& \frac{\delta^l S}{\delta \phi^A} +
(-1)^{\epsilon_M+1} \tilde{\Gamma}^M_{AK}C^K\Phi^*_M +
(i\hbar)(-1)^{\epsilon_A+\epsilon_C}
\bar{\Gamma}^C_{CB}\left(u^{-1}\right)^B_A .
\end{eqnarray}
These ``Abelianized'' transformations differ slightly from the ones of
eq. (27), but their Ward identities generate the same Schwinger-Dyson
equations, so they are equivalent.\footnote{To check this, one again
perform the ghost-antighost integrals in the identity
$0=\langle\delta[\Phi^*_AF(\phi)]\rangle$. After a number of
cancellations, one finds that the result agrees with that based on the
Abelian transformations (27).}

It should be obvious from our derivation, but we emphasize it again
here: although the
extended action itself is not invariant under the transformation (65),
the remaining term is precisely cancelled by a contribution from the
measure, {\em provided the original measure density} $\rho(\phi)$ {\em
is covariantly conserved with respect to the connection}
$\Gamma^A_{BC}$,
\beq
\rho^{-1}\frac{\delta^r \rho}{\delta\phi^A} =
(-1)^{\epsilon_S+\epsilon_A} \Gamma^S_{AS} .
\eeq
The only new property of $\Gamma^A_{BC}$ one needs in order to
demonstrate this is
\beq
(-1)^{\epsilon_S+1}\Gamma^S_{SA} +
(-1)^{\epsilon_S+\epsilon_A}\Gamma^S_{AS} = 0 ,
\eeq
which indeed can be derived directly from the definition (52). So BRST
symmetry of the path integral is again directly linked to the measure
density being covariantly conserved.

What is the Master Equation for the action in this formulation? The
goal is to generalize the solution for the extended action,
\beq
S_{ext} = S[\phi] + \Phi^*_AC^A
\eeq
to a more general function of the fields $\phi$ and the antighosts
$\Phi^*$,
\beq
S_{ext} = S^{BV}[\phi,\Phi^*] + \Phi^*_AC^A ,
\eeq
while still ensuring correct Schwinger-Dyson equations for the
classical action $S[\phi]$. The naive procedure would be to simply
replace $S$ by $S^{BV}$ in the transformation law for $\Phi^*$, and
then write down the condition that the extended action $S_{ext}$ is
invariant under this modified BRST transformation. This is sufficient
in the case of flat measures \cite{us}, but it is incorrect in the
present situation. For consistency one must demand that the Master
Equation for $S^{BV}$ does not involve the ghosts $C$. This is
achieved if we choose
\begin{eqnarray}
\delta \phi^A &=& C^A \cr
\delta C^A &=& 0 \cr
\delta\Phi^*_A &=& \frac{\delta^l S^{BV}}{\delta\phi^A} +
(-1)^{\epsilon_M+1} \tilde{\Gamma}^M_{AK}C^K\Phi^*_M + (i\hbar)
(-1)^{\epsilon_A+\epsilon_C}
\bar{\Gamma}^C_{CB}\left(u^{-1}\right)^B_A \cr && +
(-1)^{\epsilon_A\epsilon_M}\frac{\delta^r S^{BV}}{\delta
\Phi^*_B}\tilde{\Gamma}^M_{BA}\Phi^*_M .
\label{70}
\end{eqnarray}

Finding the associated Master Equation for the extended action
$S_{ext}$ is then only a question of
demanding that the BRST variation of this extended action is
cancelled by the corresponding BRST variation of the measure. With
correctly imposed boundary conditions this will still ensure that
correct Schwinger-Dyson equations are obtained for all fields involved.
In particular, before any fixings of internal gauge symmetries are
imposed, the antighosts $\Phi^*$ are simply set to zero by integrating
over the ghosts $C$. Since $S^{BV}[\phi,\Phi^*=0]\!=\!S[\phi]$, the
above follows.

Note that in this connection it is absolutely crucial that the extra
term added to the transformation law for $\delta\Phi^*_A$,
$$
(-1)^{\epsilon_A\epsilon_M}\frac{\delta^r S^{BV}}{\delta\Phi^*_B}\tilde{
\Gamma}^M_{BA}\Phi^*_M ,
$$
is independent of the ghosts $C$. Otherwise this
term could contribute to the above Ward identities, and not yield (formally)
the correct Schwinger-Dyson equations for the original theory based
on $S[\phi]$.

Reinstating the Grassmann-odd BRST transformation parameter $\mu$, and
denoting genuine variations (as opposed to the previous BRST
variations, which change statistics)\footnote{Our conventions follow
those of the Appendix in ref. \cite{us}.} by $\bar{\delta}$, we find:
\begin{eqnarray}
\bar{\delta}S_{ext} &=&  \frac{\delta^r
S^{BV}}{\delta\Phi^*_A}\frac{\delta^l S^{BV}}{\delta\phi^A}\mu
+(i\hbar)(-1)^{\epsilon_A+\epsilon_C} \frac{\delta^l
S^{BV}}{\delta\Phi^*_A} \bar{\Gamma}^C_{CB}\left(u^{-1}\right)^B_A\mu
\cr && +
(i\hbar)(-1)^{\epsilon_A+\epsilon_C}\bar{\Gamma}^C_{CB}\left(u^{-1}\right)^B_A
\mu C^A  .
\end{eqnarray}
The last term was the only part present when $S^{BV}[\phi,\Phi^*]
\!=\! S[\phi]$. It was, in that case, precisely cancelled by a similar
contribution from the measure.

In the present case we have to evaluate a new super Jacobian
associated with the BRST transformation. Using
\beq
\tilde{\Gamma}^B_{AB} = \Gamma^B_{AB} +
(-1)^{\epsilon_A(\epsilon_B+\epsilon_C
+1)}u^B_S\bar{\Gamma}^S_{CD}\left(u^{-1}\right)^D_A\left(u^{-1}\right)^C_B
,
\eeq
and a few identities based on the symmetry properties of
$\tilde{\Gamma}^A_{BC}$, one finds that the cancellation of BRST
variations of the extended action and the measure requires
\beq
\frac{\delta^r S^{BV}}{\delta\Phi^*_A}\frac{\delta^l
S^{BV}}{\delta\phi^A}\mu +
(i\hbar)(-1)^{\epsilon_A+1}\frac{\delta^r}
{\delta\phi^A}\frac{\delta^r}{\delta\Phi^*_A}S^{BV}\mu
+
(i\hbar)(-1)^{\epsilon_A+1}\rho^{-1}\left(\frac{\delta^r\rho}{\delta\phi^A}
\right)\left(\frac{\delta^r}{\delta\Phi^*_A}S^{BV}\right)\mu = 0 .
\eeq
This can be expressed compactly as
\beq
\frac{\delta^r S^{BV}}{\delta\Phi^*_A}\frac{\delta^l S^{BV}}{\delta\phi^A}
= - (i\hbar)\Delta_{\rho} S^{BV} ,
\eeq
with
\beq
\Delta_{\rho} \equiv
(-1)^{\epsilon_A+1}\rho^{-1}\frac{\delta^r}{\delta\phi^A}\left(\rho
\frac{\delta^r}{\delta\Phi^*_A}\right) .
\eeq
This $\Delta_{\rho}$ is the covariant generalization of
Batalin and Vilkovisky's
$\Delta$-operator \cite{Batalin}. Equation (74) is the generalization
of the quantum Master Equations to a theory with a non-trivial measure
of all fields $\phi$.

When $\rho\!=\!1$, there is a direct relation between
the Schwinger-Dyson BRST operator $\delta$ and the $\Delta$-operator
of Batalin and Vilkovisky \cite{us}.
In the conventional way of representing the formalism,
$(i\hbar)\Delta$ is viewed as a ``quantum correction" to
the BRST operator defined by the antibracket
\cite{Henneaux}. We now know that this particular appearance of
quantum corrections to the BRST symmetry is due to the fact that the
Batalin-Vilkovisky formalism is formulated at the
intermediate stage where the ghosts $C^A$ have been integrated out of
the path integral, but their antighost partners $\Phi^*_A$ have been kept.
Indeed, when the measure density $\rho$ is non-trivial, the same
phenomenon takes place. Just as $\Delta$ in the Master
Equation is replaced by the covariant $\Delta_{\rho}$ when the measure
density $\rho$ is non-trivial, the quantum correction to the BRST
symmetry now becomes proportional to $(i\hbar)\Delta_{\rho}$.
The identity
\beq
\int [dC] F(C^B)\exp\left[-\frac{i}{\hbar}\Phi^*_AC^A\right]
= F\left(i\hbar\frac{\delta^l}{\delta\Phi^*_B}\right)
\int [dC]\exp\left[-\frac{i}{\hbar}\Phi^*_AC^A\right]
\eeq
suffices to show this. It tells us how to correctly replace the ghosts
$C^A$ in the BRST variations when these fields have been integrated
out. The Schwinger-Dyson BRST variation of an
arbitrary functional $G[\Phi,\Phi^*]$ can, inside the functional
integral (where partial integrations are allowed), then be rewritten:
\begin{eqnarray}
&& \!\!\!\!\!\!\!\!\!\!\!\!\!\!\!\!\!\!\!\!\!\!\!\delta
G[\phi,\Phi^*] \cr &=& \frac{\delta^r G}{\delta\phi^A}C^A
+ \frac{\delta^r G}{\delta\Phi^*_A}\left\{(-1)^{\epsilon_M+1}\Gamma^M_{AK}C^K
\Phi^*_M +
(-1)^{\epsilon_A\epsilon_M+1}\frac{\delta^r S^{BV}}{\delta\Phi^*_K}
\Gamma^M_{KA}\Phi^*_M - \frac{\delta^l S^{BV}}{\delta\phi^A}\right\} \cr
&\to& \frac{\delta^r G}{\delta \phi^A}\frac{\delta^l S^{BV}}
{\delta \Phi^*_A} - \frac{\delta^r G}{\delta \Phi^*_A}
\frac{\delta^l S^{BV}}{\delta \phi^A} +
(i\hbar)\left((-1)^{\epsilon_A}\frac{\delta^r}{\delta \phi^A}
+ (-1)^{\epsilon_A\epsilon_G+\epsilon_M}\Gamma^M_{AM}\right)
\frac{\delta^r}{\delta \Phi^*_A} G.
\end{eqnarray}
The arrow has indicated where a partial integration is required.
Furthermore, by means of the identity (76) we have
\beq
\Delta_{\rho}G = (-1)^{\epsilon_A+1}\rho^{-1}\frac{\delta^r}
{\delta \phi^A} \left(\rho\frac{\delta^r G}{\delta \Phi^*_A}\right) =
\left((-1)^{\epsilon_A+1}\frac{\delta^r}{\delta \phi^A} +
(-1)^{\epsilon_A\epsilon_G+\epsilon_M+1}\Gamma^M_{AM}\right)
\frac{\delta^r}{\delta \Phi^*_A}G.
\eeq
So the equivalent of the Schwinger-Dyson BRST operator at the
intermediate stage where one has
integrated out the ghosts $C^A$ is given by (in a notation
similar to ref. \cite{Henneaux}),
\beq
\sigma_{\rho} \equiv (\,\cdot\,,S^{BV}) - i\hbar \Delta_{\rho}
\eeq
in the covariant formulation. Note that the action $S^{BV}$ is {\em not}
a generator of the correct BRST symmetry within the antibracket:
it picks up only the
classical part of the Schwinger-Dyson equations. This holds even if
the measure density $\rho$ is trivial.

As a simple check of the above manipulations, consider evaluating
the Ward identity
\beq
\langle \sigma_{\rho}\Phi^*_A F(\phi) \rangle = 0 .
\eeq
in the case where there are no internal gauge symmetries, and
$S^{BV}[\phi,\Phi^*]$ therefore can be taken to equal $S[\phi]$. It is
straightforward to confirm that the result coincides with eq. (61).
When there are internal gauge symmetries, it is important to choose
correct boundary conditions for $S^{BV}$ in order to keep the correct
Schwinger-Dyson equations as Ward identities, completely analogous to
the case of trivial measure density $\rho\!=\!1$ \cite{us}.

\vspace{1cm}

\subsection{Example: a space with torsion}

A specific example may be instructive at this point. Suppose we are given
a space of fields, a manifold (and let us for simplicity choose it to
be entirely bosonic), and suppose that we are provided with a natural
connection $\Gamma^A_{BC}$ on this manifold. Can we set up a suitable
quantization procedure for a theory of action $S_{ext}$ defined on such
a manifold? Since nothing has been assumed about the symmetry properties
of the connection $\Gamma^A_{BC}$, we split it up into its symmetric
and antisymmetric combinations (with respect to the lower indices):
\beq
\Gamma^A_{BC} = \frac{1}{2}(\Gamma^A_{BC} + \Gamma^A_{CB}) +
S^{\,\,\cdot\,\,\cdot\,A}_{BC} ,
\label{81}
\eeq
where the torsion tensor $S^{\,\,\cdot\,\,\cdot\,A}_{BC}$ is defined to
the antisymmetric part of the connection:
\beq
S^{\,\,\cdot\,\,\cdot\,A}_{BC} \equiv \frac{1}{2}\left[\Gamma^A_{BC}
- \Gamma^A_{CB}\right] .
\eeq

We shall now seek the appropriate Schwinger-Dyson BRST algebra for field
spaces with torsion. The only modification in comparison with the (bosonic)
curved, but torsionless, case is that the connection $\Gamma^A_{BC}$ is
no longer symmetric in the lower indices. As far as the action is
concerned, BRST invariance hinges in the bosonic case crucially on this
symmetry. This means that if we blindly substitute the connection (\ref{81})
into the conventional Schwinger-Dyson BRST transformations (\ref{27}), the
action will no longer be invariant. We can cure for this by taking only
a specific symmetric combination $\Lambda^A_{BC}$ defined by
\beq
\Lambda^A_{BC} \equiv \frac{1}{2}\left[\Gamma^A_{BC}+\Gamma^A_{CB}\right]
+ \left[S^{A\,\,\cdot\,\,\cdot}_{\,\,\,BC}+
S^{A\,\,\cdot\,\,\cdot}_{\,\,\,CB}\right] .
\label{83}
\eeq

The action is now guaranteed to be invariant under the BRST transformations
\begin{eqnarray}
\delta\phi^A &=& c^A \cr
\delta c^A &=& 0 \cr
\delta\phi^*_A &=& \Lambda^C_{BA}c^B\phi^*_C - \frac{\delta S}
{\delta\phi^A} ,
\label{84}
\end{eqnarray}
but we still need to check that the functional measure is invariant as
well. This is the case if
\beq
\frac{\delta \rho}{\delta\phi^A} - \Lambda^B_{AB}\rho = 0,
\eeq
which, on account of eq. (\ref{83}), is equivalent to
\beq
\frac{\delta \rho}{\delta\phi^A} - \Gamma^B_{AB}\rho = 0 ,
\label{86}
\eeq
that is, precisely the condition that $\rho$ is covariantly conserved
with respect to the connection $\Gamma^A_{BC}$.
An invariant measure defined through this criterion will thus satisfy
all requirements.

To make these considerations even more concrete, let us study explicitly
a particular case: a Riemann-Cartan space defined by a metric-preserving
connection $\Gamma^A_{BC}$, which can be written in terms of the Christoffel
symbol and the so-called {\em contortion tensor} $K^{\,\,\cdot\,\,\cdot A}_
{BC}$ in the following way \cite{Hehl}:
\beq
\Gamma^A_{BC} = \left\{^{\,A}_{BC}\right\} - K^{\,\,\cdot\,\,\cdot A}_{BC}.
\eeq
The tensors $K$ and $S$ are related to each other:
\beq
K^{\,\,\cdot\,\,\cdot A}_{BC} = -S^{\,\,\cdot\,\,\cdot A}_{BC} +
S^{\,\,\cdot\,
A\,\,\cdot}_{B\,\cdot\,\,C} - S^{A\,\,\cdot\,\,\cdot}_{\,\,\cdot \,BC} .
\eeq

An explicit expression for an invariant measure can be found by solving the
constraint that $\rho$ be covariantly conserved with respect to $\Gamma$,
eq. (\ref{86}). An integrability condition is that a trace of the torsion
tensor can be derived from a scalar potential, $S^A_{AB} = \partial
\Theta/\partial\phi^B$, in terms of which (see, e.g., ref. \cite{Saa}):
\beq
\rho(\phi) = e^{2\Theta(\phi)}\sqrt{g(\phi)} .
\eeq

We now add to the action the same $\phi^*_Ac^A$-term as before.
The BRST Ward identities $0 = \langle \delta (\phi^*_A F[\phi])\rangle$
correspond, after integrating out the ghost-antighost pair $c^A,\phi^*_A$,
to the Schwinger-Dyson equations for a theory of action $S$ and measure
density $\rho(\phi)$ satisfying eq. (\ref{86}). The BRST symmetry (\ref{84}) is
therefore
the sought-for Schwinger-Dyson BRST symmetry for this case.

\vspace{1cm}

\section{Master Equations for other symmetries}

While Schwinger-Dyson equations are the most general identities of any
quantum field theory, there are of course interesting subsets that can
play very important r\^{o}les. The Ward identities of ordinary gauge
symmetries, chiral Ward identities, conformal Ward identities, etc.,
are all examples that illustrate the importance of having exact
identities which one can demand must be satisfied by a quantum theory.
Thus, while Schwinger-Dyson equations can serve to define the full
Lagrangian quantum theory \cite{us}, there may be less general
equations that one wishes to impose in the process of quantization.
Having realized how to derive the Batalin-Vilkovisky Lagrangian
formalism \cite{Batalin} from the Schwinger-Dyson BRST symmetry
\cite{us0}, it is of course not a big step to generalize this to an
arbitrary class of transformations. This is the subject we turn to next.

The idea is to promote an arbitrary symmetry of action or of measure
(or of both) into a BRST symmetry. The original symmetry may be local
or global, and it is irrelevant if it is spontaneously broken. In
fact, the symmetry may even be anomalous in the sense that it is
broken by quantum corrections. To make the picture complete, we can
even demand that the BRST symmetry is based on transformations of the
original fields that leave neither the action nor the measure, {\em
nor even the combination of the two}, invariant. That this is
possible, is due to the enormous freedom we have available in defining
the Lagrangian path integral.

\subsection{Non-invariant measures}

Suppose we are presented with a partition function
\beq
{\cal Z} = \int [d\phi]\rho(\phi)e^{\frac{i}{\hbar}S[\phi]} .
\eeq
In the previous section we discussed the set of transformations (\ref{37})
that left the functional measure $[d\phi]\rho(\phi)$ invariant. Here
we want to be more general, and consider this theory in the light of
{\em arbitrary} transformations $\phi^A = g^A(\phi',a)$, independently
of whether they leave the measure invariant or not. What is the
modification? When we perform the transformation that takes us from
$\phi$-variables to $\phi'$-variables, the measure changes
due to the Jacobian:
\beq
\int [d\phi]\rho(\phi) \to \int [d\phi']\rho(\phi',a){\mbox{\rm sdet}}(M) ,
\label{91}
\eeq
where the matrix $M^A_B$ is as defined in eq. (\ref{23}), and where
$\rho(\phi',a) = \rho(\phi(\phi',a))$. The fact that the measure
changes due to the field transformation does not affect the new gauge
symmetry between the transformed fields $\phi'$ and and the collective
fields $a$. Consider the Abelian formulation. We here integrate $a$
over the flat measure, which is invariant under the gauge
transformation $\delta a^i\!=\!\varepsilon^i$. The full measure
(\ref{91}) for
$\phi'$ is now invariant under the gauge transformation corresponding
to (\ref{22}).
This is obvious from its construction (the measure was
originally only a function of the fields $\phi^A$, and these are gauge
invariant by themselves), but it is far from obvious when one
considers the definition (\ref{91}) of the $\phi'$-measure.
It is therefore worthwhile to check the invariance. We use the fact
that

\beq
\delta[{\mbox{\rm sdet}}(M)] = {\mbox{\rm sdet}}(M){\mbox{\rm
sTr}}[M^{-1}\delta M] .
\eeq
The measure density $\rho(\phi',a)\!=\!\rho(\phi)$ is explicitly
invariant by itself, since it is only a function of $\phi$. Finally,
from the $[d\phi']$ part we get an extra term
$$
{\mbox{\rm sTr}}\left[\frac{\delta^r \delta\phi'}{\delta\phi'}\right].
$$
It follows from eq. (\ref{23}) that
\begin{eqnarray}
\frac{\delta^r \delta \phi'^A}{\delta \phi'^C} &=& -\left(M^{-1}\right)^A_B
\frac{\delta^r M^B_C}{\delta a^i}\delta a^i +
(-1)^{\epsilon_C\epsilon_B}
\frac{\delta^r\left(M^{-1}\right)^A_B}{\delta\phi'^C} M^B_D\delta\phi'^D
\cr &=& -\left(M^{-1}\right)^A_B
\frac{\delta^r M^B_C}{\delta a^i}\delta a^i - \left(M^{-1}\right)^A_B
\frac{\delta^r M^B_C}{\delta\phi'^D}\delta\phi'^D \cr
&=& - \left(M^{-1}\right)^A_B\delta M^B_C ,
\end{eqnarray}
which shows that the Jacobian from $[d\phi']$ is cancelled by the
change in sdet$(M)$. So the theory defined formally by
\beq
{\cal Z} = \int [da][d\phi']\rho(\phi',a){\mbox{\rm
sdet}}[M(\phi',a)]e^{\frac{i}{\hbar} S[\phi',a]}
\eeq
has the gauge symmetry (\ref{37}). The measure and the action are separately
invariant. The transformations $g^A(\phi',a)$ can hence be chosen
completely arbitrary; the path integral enlarged with the help of the
collective fields $a$ is still defining a gauge invariant theory. We
only need to include a possible Jacobian factor sdet$(M)$, as shown above.

If the transformations are chosen arbitrarily, how can they contain
any physical information? They can because they probe the response of
the path integral to a reparametrization. If we perform functional
averages, these probes of the path integral give rise to identities,
in general a subset of the complete set of Schwinger-Dyson equations.
This is no more surprising than the previous case of invariant measures:
also here the only information concerns properties of the measure,
and is completely independent of the action $S$ under consideration.

The procedure is now the same as before:
when we integrate out the collective fields
$a$ after having fixed an appropriate gauge, the left-over BRST
symmetry will give rise to non-trivial Ward identities. Of course,
these Ward
identities are not new. Just as in the case of Schwinger-Dyson
equations, they can be derived straightforwardly by manipulations
directly at the path-integral level. However, as with
Schwinger-Dyson equations, one advantage of the corresponding BRST
formulation is that the identities can be
{\em imposed} on the path integral by means of a non-trivial condition
on the quantum action -- a generalized Master Equation. One can then
select the whole class of actions that will yield the same identities,
for all the fields involved.

As before, we fix $a^i\!=\!0$ by adding to the action a term of the
form $-\delta[\phi^*_A a^A]$. The BRST operator is the same as in eq.
(\ref{22}), since the
gauge symmetry is unaffected by the presence of the extra sdet$(M)$
term in the measure. Integrating out $B_i$ and $a^i$ we do, however,
get a slight modification of the residual BRST transformation law:
\begin{eqnarray}
\delta\phi^A &=&-u^A_i(\phi)c^i \cr
\delta c^i &=& 0 \cr
\delta\phi^*_i &=&(-1)^{\epsilon_i}\left[\frac{\delta^r
S}{\delta \phi^A}u^A_i(\phi) - (i\hbar)
\left\{\frac{\delta^r\ln\rho}{\delta\phi^A}u^A_i +
\left.\frac{\delta^r{\ln\mbox{\rm
sdet}}(M)}{\delta a^i}\right|_{a=0}\right\}\right] .
\end{eqnarray}
In the case of a measure invariant under the substitution
$\phi^A\!=\!g^A(\phi',a)$, the last two terms cancel each other.

We find the relevant BRST Ward identities as before by evaluating the
ghost--antighost expectation value
$0\!=\!\langle\delta[\phi^*_iF(\phi)] \rangle$. This gives
\beq
\left\langle(-1)^{\epsilon_i(\epsilon_A+1)}u^A_i\left[\frac{\delta^l
F}{\delta\phi^A} + \left(\frac{i}{\hbar}\right)\frac{\delta^l
S}{\delta\phi^A}F\right]\right\rangle =
\left\langle(-1)^{\epsilon_i+1}\left[
\frac{\delta^r\ln\rho}{\delta\phi^A}u^A_i + \left.\frac{\delta^r\ln{\mbox{\rm
sdet}}(M)} {\delta a^i}\right|_{a=0}\right]\right\rangle .
\label{106}
\eeq

While the identities (\ref{106}) are not as general as the full set of
Schwinger-Dyson equations, there are many examples in field theory
where they play an important r\^{o}le. Typical cases may involve anomalous
Ward identities, where the measure can be formally invariant under a group
of transformations, but where this symmetry is broken by the
ultraviolet regulator such as a set Pauli-Villars fields. If these
fields are integrated out of the path integral, they will provide a
non-invariant measure for the original variables.
The right hand side of eq. (\ref{106}) then provides
the violation of the naive Ward identity.

As with the case of invariant measures, the present Abelian formulation
suffers from the problem that the BRST charge is not nilpotent when acting
on the fields $\phi$. It fortunately
takes little work to see that the new measure
$$
\int [da]_E[d\phi']_E\rho(\phi',a){\mbox{\rm sdet}}\lambda(a){\mbox{\rm
sdet}}M(\phi',a)
$$
is invariant under the corresponding non-Abelian gauge transformation
(\ref{37})
as well. In order to compute the super Jacobian of the transformation,
we first evaluate
\beq
\frac{\delta^r \delta\phi'^A}{\delta\phi'^B} = (-1)^{\epsilon_B\epsilon_i}
\frac{\delta^r u^A_i}{\delta\phi'^B}\varepsilon^i .
\eeq
Next, differentiating the identity (\ref{25}) with respect to $\phi'$, we find:
\beq
\frac{\delta^r u^D_j}{\delta\phi'^C} = \left(M^{-1}\right)^D_A\left[
(-1)^{\epsilon_i\epsilon_C}\frac{\delta^r M^A_C}{\delta a^i}\nu^i_j
- (-1)^{\epsilon_C(\epsilon_B+\epsilon_j)}\frac{\delta^r M^A_B}{\delta\phi'^C}
u^B_j\right] ,
\eeq
which means that
\begin{eqnarray}
\frac{\delta^r \delta\phi'^A}{\delta\phi'^B} &=& \left(M^{-1}\right)^A_C
\left[\frac{\delta^r M^C_B}{\delta a^j}\nu^j_i - (-1)^{\epsilon_B\epsilon_D}
\frac{\delta^r M^C_D}{\delta\phi'^B}u^D_i\right]\varepsilon^i \cr
&=& \left(M^{-1}\right)^A_C\left[\frac{\delta^r M^C_B}{\delta a^j}\nu^j_i
- \frac{\delta^r M^C_B}{\delta\phi'^D}u^D_i\right]\varepsilon^i \cr
&=& -\left(M^{-1}\right)^A_C\delta M^C_B .
\end{eqnarray}
As in the Abelian case, the Jacobian from $[d\phi']$ is therefore
precisely cancelled by the change in sdet$(M)$: the new measure is invariant
under the non-Abelian gauge symmetry.

At the level where all fields are kept, the corresponding non-Abelian
BRST transformations obviously coincide with those of eq. (\ref{37}).
Differences
only show up when we gauge-fix the symmetry (by, say, adding the conventional
$-\delta\{\phi^*_ia^i\}$ term to the action) and then integrate out both the
collective fields $a^i$ and the
Nakanishi-Lautrup fields $B_i$. Using the same technique as in the section
above, we find that only the transformation law for $\phi^*_i$ is
modified:
\begin{eqnarray}
\delta \phi^*_i =&& (-1)^{\epsilon_i}\frac{\delta^r S}{\delta\phi^A}
u^A_i(\phi)+ (-1)^{\epsilon_i\epsilon_j}\phi^*_k\bar{\Gamma}^k_{ji}c^j \cr
&& +(i\hbar)\left[(-1)^{\epsilon_i+\epsilon_j}\bar{\Gamma}^j_{ji} +
(-1)^{\epsilon_i+1}\left\{\frac{\delta^r\ln\rho}{\delta\phi^A}u^A_i(\phi)
+ N_i\right\}\right] ,
\label{110}
\end{eqnarray}
where we have defined
\beq
N_i \equiv \left.\frac{\delta^r\ln{\mbox{\rm sdet}}(M)}{\delta a^i}
\right|_{a=0} = (-1)^{\epsilon_A}\left.\frac{\delta^r M^A_A}{\delta
a^i}\right|_{a=0} .
\eeq

\subsection{The corresponding Master Equation}

Although the action $S_{ext}$ is not invariant under (\ref{110}), we also here
have the
situation that its BRST variation $\delta S_{ext}$ is cancelled by a
contribution
from the measure. So far we have assumed that $S$ is a function of the fields
$\phi^A$ only. We can easily generalize this to an arbitrary action $S_{ext}$
that depends on
both the fields $\phi^A$ and, say, the antighosts $\phi^*_A$\footnote{To
ensure that this extended action $S_{ext}$ has zero ghost number, this
of course requires the presence of fields $\phi^A$ with non-zero ghost
number. Such ghost fields can either already be
inherent in the formalism (required by the subsequent gauge fixing of
internal symmetries),
or they can be added by hand.}. As before, depending on the manner in which
we generalize the action, we are forced to modify the BRST transformation
laws as well. Since we wish to preserve nilpotency of the BRST charge $Q$
when acting on the fundamental fields $\phi^A$, the only possibility is to
modify the transformation law for the antighosts $\phi^*$.

Let us again restrict ourselves to the generalization
\beq
S \to S_{ext} = S^{BV}[\phi,\phi^*] + \phi^*_ic^i .
\eeq
Using the ansatz
\begin{eqnarray}
\delta \phi^*_i =&& (-1)^{\epsilon_i}\frac{\delta^r S^{BV}}{\delta\phi^A}
u^A_i(\phi)+ (-1)^{\epsilon_i\epsilon_j}\phi^*_k\bar{\Gamma}^k_{ji}c^j
+ M_i \cr
&& +(i\hbar)\left[(-1)^{\epsilon_i+\epsilon_j}\bar{\Gamma}^j_{ji} +
(-1)^{\epsilon_i+1}\left\{\frac{\delta^r\ln\rho}{\delta\phi^A}u^A_i(\phi)
+ N_i\right\}\right] ,
\end{eqnarray}
we determine the extra term $M_i$ from the requirement that the corresponding
Master Equation for $S^{BV}$ is independent of the ghosts $c^i$ (since this
would contradict the assumption that $S^{BV}$ depends only upon $\phi^A$ and
$\phi^*_i$). As the associated Master Equation will contain both classical
and quantum (proportional to $\hbar$) parts, the possibility of finding a
consistent solution for $M_i$ is not at all guaranteed. At the classical
level, we find that the two terms in $\bar{\delta}S_{ext}$ which involve
two factors of ghosts $c$ cancel each other automatically, independently
of the choice of $M_i$. Staying at the classical level, terms involving
only one factor of $c$ cancel if
\beq
M_i = (-1)^{\epsilon_i(\epsilon_j+1)}\frac{\delta^r S^{BV}}{\delta
\phi^*_j}\phi^*_k\bar{\Gamma}^k_{ij} .
\label{114}
\eeq

This also immediately gives us the classical Master Equation:
\beq
(-1)^{\epsilon_i}\frac{\delta^r S^{BV}}{\delta\phi^*_i}\frac{\delta^r S^{BV}}
{\delta\phi^A}u^A_i(\phi) + \frac{1}{2}(-1)^{\epsilon_i(\epsilon_j+1)}
\frac{\delta^r S^{BV}}{\delta\phi^*_i}\frac{\delta^r S^{BV}}{\delta\phi^*_j}
\phi^*_k c^k_{ij} = 0.
\label{115}
\eeq
Note that the usual antibracket does not appear here (there are not,
in general, an
equal number of ``coordinates'' $\phi^A$ and ``momenta'' $\phi^*_i$, so this
was ruled out from the beginning). And a new term, proportional to the
structure coefficients of the group of transformations has emerged.

Being described purely at the classical
level, the above Master Equation of course makes no reference to the
functional measure,
and to whether this measure is invariant under the field transformation
$\phi^A = g^A(\phi',a)$ or not. This means that the same Master Equation
should emerge even in the more conventional case described in section 2
where we considered the case of a one-to-one matching between fields
$\phi^A$ and collective fields $a^A$. When the transformation matrix
$u^A_B(\phi)$ was invertible, we could show that the associated classical
Master Equation was nothing but the usual antibracket relation
\beq
(S^{BV},S^{BV}) = 0 .
\eeq
On the surface, this would seem to contradict the derivation presented
here. The case of a one-to-one matching of degrees of freedom between
$\phi^A$ and $a^A$, and an invertible $u^A_B(\phi)$, is but a special
case of the above more general considerations. Fortunately, there is
no contradiction. The difference between
these two alternative descriptions lies, in the special case referred
to in section 2
in the choice of antighosts $\phi^*$. When $u^A_B$ is invertible,
we can define new ghosts $C^A$ and new antighosts $\Phi^*_A$ according
to eq. (\ref{62}). This does not affect the action
\beq
S_{ext} = S[\phi] + \phi^*_Ac^A ,
\eeq
which is invariant under such a substitution. But when we next generalize the
extended action to include an antighost-dependence in the main part, the
Master Equation for
\beq
S_{ext} = S[\phi,\phi^*] + \phi^*_Ac^A
\eeq
will of course differ from
\beq
S_{ext} = S[\phi,\Phi^*] + \Phi^*_AC^A
\eeq
due to the implicit $\phi$-dependence in $\Phi^*$. In fact, the term needed
to provide the same solution (in terms of the same variables) for these
two equations is precisely the ``commutator'' term in eq. (\ref{115}) above.

The fact that $M_i$ turns out to involve only one power of $\phi^*$, and
none of the ghosts $c$, is crucial for the consistency of this procedure.
This way the Ward identities for $S_{ext}$ will, formally, coincide with
the Schwinger-Dyson equations for the original action $S[\phi]$ after
integrating over the ghosts $c$ and antighosts $\phi^*$. This however, is
not the only consistency check. Although the choice (\ref{114})
guarantees the absence of terms involving the ghosts $c$ in the classical
Master Equation for $S^{BV}$, nothing would in principle prevent
$c$-dependent terms at the quantum level. This again would spoil the
consistency of this procedure for determining the transformation law
of $\phi^*$. Thus one has to find the full quantum Master Equation
and check that it indeed is $c$-independent
before one is sure of having a consistent formulation.

It still remains to be checked whether the additional term (\ref{114})
also suffices to guarantee that the quantum mechanical Master
Equation for $S^{BV}$ is independent of the ghosts $c$. This is
indded the case, and we then finally have the complete quantum
Master Equation.
{}From the variation of the action we get:
\begin{eqnarray}
\delta S_{ext}=\frac{\delta^r S^{BV}}{\delta\phi^*_i}
\frac{\delta^r S^{BV}}{\delta\phi^*_j}
(-1)^{\epsilon_i\epsilon_j
+\epsilon_i}\phi^*_k\bar{\Gamma}^k_{ij} +
\frac{\delta^r S^{BV}}{\delta\phi^*_i}[(-1)^{\epsilon_i}
\frac{\delta^r S^{BV}}
{\delta\phi^B}u^B_i-i\hbar(-1)^{\epsilon_i+\epsilon_j}\bar{\Gamma}^j_{ji}]
\end{eqnarray}

The Jacobian contributes with:
\begin{eqnarray}
J=-i\hbar [\frac{\delta^r}{\delta\phi^*_i}[\frac{\delta^r S^{BV}}
{\delta\phi^B} u^B_i]+\frac{\delta^r}{\delta\phi^*_i}
[\frac{\delta^r S^{BV}}{\delta\phi^*_j} (-1)^{\epsilon_i\epsilon_j}\phi^*_k
\bar\Gamma^k_{ij}]] ~.
\end{eqnarray}

We get finally the following quantum Master Equation:
\begin{eqnarray}
0 ~=~ \delta S_{ext}+ J=
\frac{\delta^r S^{BV}}{\delta\phi^*_i}
\frac{\delta^r S^{BV}}{\delta\phi^*_j}
(-1)^{\epsilon_i\epsilon_j
+\epsilon_i}\phi*_k\bar{\Gamma}^k_{ij} +
\frac{\delta^r S^{BV}}{\delta\phi^*_i}[(-1)^{\epsilon_i}
\frac{\delta^r S^{BV}}
{\delta\phi^B}u^B_i] \nonumber \\
-i\hbar [(-1)^{(\epsilon_i+1)(\epsilon_B+1)}
\frac{\delta^r}{\delta\phi^*_i}(\frac{\delta^r S^{BV}}
{\delta\phi^B}) u^B_i+(-1)^{\epsilon_j}
\frac{\delta^r}{\delta\phi^*_i}
(\frac{\delta^r S^{BV}}{\delta\phi^*_j})
\phi^*_k\bar{\Gamma}^k_{ij}] ~.
\label{113}
\end{eqnarray}

When does the more general Master Equation (\ref{115}) have non-trivial
solutions?\footnote{A $\phi^*$-independent action $S^{BV}[\phi,\phi^*] \!=\!
S[\phi]$ is of course always a solution.} Let us first recapitulate some
basic facts about the solutions to the conventional classical Master Equation
\beq
\frac{1}{2}(S^{BV},S^{BV}) = \frac{\delta^r S^{BV}}{\delta\phi^A}
\frac{\delta^l S^{BV}}{\delta\phi^*_A} = 0 .
\label{120}
\eeq
Since the most fundamental boundary condition is $S^{BV}[\phi,\phi^*\!\!
=\!\!0] \!=\! S[\phi]$, where $S$ is the classical action, it is natural
to assume that $S^{BV}$ will have an expansion in terms of antighosts
$\phi^*$:
\beq
S^{BV}[\phi,\phi^*] = S[\phi] + \phi^*_A{\cal{R}}^A(\phi) + \ldots
\eeq
Inserting this into the classical Master Equation (\ref{120}), and keeping only
the first order in $\phi^*$, leads to the equation
\beq
\left(\frac{\delta^r S}{\delta\phi^A} + \phi^*_B\frac{\delta^r{\cal{R}}^B}
{\delta\phi^A}\right){\cal{R}}^A = 0 .
\eeq
This can only be satisfied if simultaneously
\beq
\frac{\delta^r S}{\delta\phi^A}{\cal{R}}^A = 0 ,
\eeq
and
\beq
\phi^*_B\frac{\delta^r{\cal{R}}^B}{\delta\phi^A}{\cal{R}}^A = 0 .
\eeq
The first of these two equations says that the classical action must be
invariant with respect to ``internal'' BRST transformations $\cal{R}$:
\beq
\delta \phi^A = {\cal{R}}^A ,
\eeq
while the second of these two equations expresses the condition that this
symmetry be nilpotent:
\beq
\delta^2\phi^B = \delta{\cal{R}}^B = \frac{\delta^r{\cal{R}}^B}
{\delta\phi^A}{\cal{R}}^A = 0 .
\eeq

To lowest order in an expansion in antighosts $\phi^*$, there is thus, with
the boundary condition $S^{BV}[\phi,0] \!=\! S[\phi]$ imposed, precisely
a non-trivial solution to the classical Master
Equation whenever the classical action is invariant under a nilpotent BRST
symmetry.

Consider now the more general classical Master Equation (\ref{115}). We again
impose the condition $S^{BV}[\phi,0] \!=\! S[\phi]$ (because otherwise we
do not recover the correct Ward Identities from $S^{BV}$), and hence
assume an expansion
\beq
S^{BV}[\phi,\phi^*] = S[\phi] + \phi^*_ir^i(\phi) + \ldots
\label{127}
\eeq
Retaining only terms up to one power of antighosts $\phi^*$ means that
the classical Master Equation turns into
\beq
-\left(\frac{\delta^r S}{\delta\phi^A} + \phi^*_i\frac{\delta^rr^i}
{\delta\phi^A}\right)u^A_j(\phi)r^j(\phi) + \frac{1}{2}(-1)^{\epsilon_j(
\epsilon_i+1)}r^i(\phi)r^j(\phi)\phi^*_k c^k_{ij} = 0 .
\eeq

Also here this equation is actually equivalent to two separate requirements
(since different powers of $\phi^*$ must cancel individually):
\beq
\frac{\delta^r S}{\delta\phi^A}u^A_ir^i = 0 ,
\eeq
and
\beq
-\phi^*_i\frac{\delta^r r^i}{\delta\phi^A}u^A_jr^j + \frac{1}{2}
(-1)^{\epsilon_j(\epsilon_i+1)}r^ir^j\phi^*_k c^k_{ij} = 0 .
\label{130}
\eeq
The first implies that the classical action must be invariant under a
certain internal BRST symmetry,
\beq
\delta \phi^A = u^A_ir^i ,
\label{131}
\eeq
but the second condition is not at first sight related to nilpotency of
this BRST transformation. Consider, however, the condition $\delta^2
\phi^A \!=\! 0$:
\beq
\delta^2\phi^A = \delta[u^A_ir^i] = u^A_i\frac{\delta^r r^i}{\delta\phi^B}
u^B_jr^j + (-1)^{\epsilon_i}\frac{\delta^r u^A_i}{\delta\phi^B}
u^B_jr^jr^i = 0 .
\label{132}
\eeq
Next, multiplying the identity (\ref{132}) from the right by $(-1)^{\epsilon_i}
r^jr^i$, and summing over $i$ and $j$ gives
\beq
(-1)^{\epsilon_i}\frac{\delta^r u^A_i}{\delta\phi^B}u^B_jr^jr^i = -
\frac{1}{2}(-1)^{\epsilon_i}u^A_k c^k_{ij}r^jr^i ,
\eeq
which means that nilpotency of the transformation (\ref{131}) can be expressed
as
\beq
u^A_i\frac{\delta^r r^i}{\delta\phi^B}u^B_jr^j - \frac{1}{2}(-1)^
{\epsilon_i}u^A_k c^k_{ij}r^jr^i = 0 .
\eeq
It is now straightforward to verify that the condition (\ref{130}) is precisely
equivalent to the requirement that the internal BRST symmetry
$\delta\phi^A = u^A_ir^i$ is nilpotent. So as in the case of the conventional
classical Master Equation, there is also here a direct link between having
non-trivial solutions of the Master Equation to lowest order in a
$\phi^*$-expansion, and having an internal nilpotent BRST symmetry
of the action $S$. Furthermore, we see that the unusual commutator-term
in the new Master Equation (\ref{115}) is there to guarantee nilpotency of the
internal BRST symmetry, once a solution to the equation has been found.

Only in the special case of an invertible $u^A_i$ does the argument also
run in the inverse direction. There, given an internal nilpotent BRST
symmetry, one can immediately write down the lowest-order solution to
the Master Equation (since $r^i$ in that case is given explicitly in terms
of the internal BRST symmetry). But in general, all one can infer is that
if there is a non-trivial solution (\ref{127}) to the new
Master Equation (\ref{115}) (and if one knows the set of transformations $g^A$)
then the classical action will be BRST invariant with respect to
the nilpotent transformations given above.

What, then, is the advantage of having a generalized Master Equation of
the kind (\ref{115}) available? In the conventional antibracket formalism the
sole
purpose of the Master Equation is to provide an extended action with the
same physics as that of the original classical action (and, in its covariant
generalizations, of the functional measure of these fields). But the
extended action reduces to the original action if there are no internal
symmetries to be fixed. So the real purpose of the standard antibracket
formalism is to provide a systematic approach to the gauge fixing of internal
symmetries. In the language of ref. \cite{us} it is obvious
why gauge fixing is so conveniently performed at the level of the extended
action, rather than in the manner of the conventional Lagrangian BRST
technique. This is because the nilpotent Schwinger-Dyson BRST symmetry
is far more simple than that which can be encountered in the internal
symmetries (including open algebras, reducible gauge symmetries etc.),
and gauge fixing simply consists in the trivial addition of the
(Schwinger-Dyson) BRST variation of a certain gauge fermion,
\beq
\delta\Psi(\phi) = \frac{\delta^r\Psi}{\delta\phi^A}c^a ,
\eeq
to the extended action. Of course, when the classical action we are
considering indeed has an internal BRST symmetry of the factorizable form
(\ref{131}),
we can do the same kind of gauge fixing in the present formalism.

In performing that gauge fixing, it must be kept in mind that the BRST
symmetry we use is the one on which the quantization itself is based, in
this case the symmetry (\ref{70}). It does {\em not} make any reference to
possible internal BRST symmetries (instead, these surface automatically
when we solve the Master Equation, and impose the proper boundary
conditions). We thus add, for a certain gauge fermion $\Psi(\phi)$,
\beq
\delta\Psi(\phi) = \frac{\delta^r\Psi}{\delta\phi^A}u^A_i(\phi)c^i
\eeq
to the extended action. After integrating out the ghosts $c^i$, the
partition function can be written
\beq
{\cal{Z}} = \int [d\phi][d\phi^*]\delta\left(\phi^*_i - \frac{\delta^r
\Psi}{\delta\phi^A}u^A_i\right)\exp\left[\frac{i}{\hbar}S^{BV}\right] ,
\eeq
where $S^{BV}$ is a solution to the Master Equation (\ref{115}). This is
essentially in the form of the Batalin-Vilkovisky prescription
\cite{Batalin}, (although, of course, the antighosts $\phi^*_i$ can no
longer be viewed as ``antifields'' of the fields $\phi^A$)
at least in the sense that the integration over the antighosts $\phi^*_i$
is trivial (due to the $\delta$-function arising from integrating over
the ghosts $c^i$). The substitution
\beq
\phi^*_i \to \frac{\delta^r\Psi}{\delta\phi^A}u^A_i
\eeq
in the solution to the Master Equation is thus what constitutes gauge
fixing in this case. Indeed, in the action the result of such a
substitution is
\beq
S[\phi] \to S[\phi] + \frac{\delta^r\Psi}{\delta\phi^A}u^A_ir^i + \ldots ,
\eeq
which, for a closed and irreducible internal algebra, would be the result
of gauge fixing directly the internal symmetry $\delta \phi^A = u^A_ir^i$
(modulo quantum corrections). This is reassuring, because
in the particular case of an invertible $u^A_i$ the quantization
prescription based on the Master Equation (\ref{115}) is equivalent to the
one based on the conventional equation, and in this particular limit
the two results should of course coincide.

\section{A New Bracket}

The more general Master Equation derived in subsection 4.2 hints at the
existence of a new bracket structure which is more general than that of
the usual antibracket. Since we in general will not have a one-to-one
matching of fields $\phi^A$ and antighosts (``antifields'') $\phi^*_i$,
it is obvious that these fields cannot in general
be canonical within the new
bracket. However, just as the conventional antibracket can be viewed as
a Grassmann-odd bracket based on a Heisenberg algebra between fields
and antifields, a possible
generalization of the antibracket can be based on a more general algebra.
Indeed, as we shall now show, such a generalization is at the heart of
the more general Master Equation formalism derived above.

\subsection{The more general ``Quantum BRST Operator''}

To start, let us consider the by now simple problem of deriving the analogue
of the ``quantum BRST operator'' associated with the more general BRST
symmetry (\ref{51}). We again wish to see how the BRST symmetry can be
represented
when the ghosts $c^i$ have been integrated out of the path integral. We
use the identity
\beq
\int [dc] F(c^j)e^{\frac{i}{\hbar}\phi^*_ic^i} = F\left(-i\hbar
\frac{\delta^l}{\delta\phi^*_j}\right)\int [dc]e^{\frac{i}{\hbar}\phi^*_ic^i}
,
\eeq
and consider the BRST-variation of an arbitrary Green function $G$
depending only upon fields $\phi^A$ and antighosts $\phi^*_i$. For simplicity,
consider the particular case when the functional measure is invariant under
the transformation $\phi^A \to g^A(\phi',a)$. Using similar
manipulations as in section 2.2 (most notably, a partial integration inside
the path integral), we find that
\beq
\delta G(\phi,\phi^*) \to \bar{\sigma} G(\phi,\phi^*) ,
\eeq
where
\beq
\bar{\sigma}G \equiv [G,S^{BV}] - i\hbar\bar{\Delta}G .
\label{156}
\eeq
Here $\bar{\Delta}$ is defined by
\beq
\bar{\Delta}G \equiv (-1)^{\epsilon_i}\left[\frac{\delta^r}{\delta\phi^A}
\frac{\delta^r}{\delta\phi^*_i}G\right]u^A_i + \frac{1}{2}(-1)^{\epsilon_i+1}
\left[\frac{\delta^r}{\delta\phi^*_j}\frac{\delta^r}{\delta\phi^*_i}G\right]
\phi^*_kc^k_{ji} ,
\eeq
and $[\cdot,\cdot]$ denotes a new Grassmann-odd bracket:
\beq
[F,G] \equiv (-1)^{\epsilon_i(\epsilon_A+1)}\frac{\delta^r F}
{\delta\phi^*_i}u^A_i\frac{\delta^l G}{\delta\phi^A} - \frac{\delta^r F}
{\delta\phi^A}u^A_i\frac{\delta^l G}{\delta\phi^*_i} + \frac{\delta^r F}
{\delta\phi^*_i}\phi^*_kc^k_{ij}\frac{\delta^l G}{\delta\phi^*_j} .
\eeq

\subsection{Properties of the New Bracket}

The bracket structure $[\cdot,\cdot]$ defined above is what enters naturally
when one considers the BRST operator at the level where the ghosts $c^i$
have been integrated out, but their antighost partners $\phi^*_i$ have been
kept. It is not obvious that such a bracket structure is of relevance
beyond that stage. But the new bracket turns out to possess a number of
useful properties that elevates it to a somewhat higher status. These
properties are all shared with the conventional antibracket.

First, one easily verifies that the new bracket indeed is statistics-changing
in the sense that $\epsilon([F,G]) = \epsilon(F)+\epsilon(G)+1$. It
satisfies an exchange relation of the kind
\beq
[F,G] = (-1)^{\epsilon_F\epsilon_G+\epsilon_F+\epsilon_G}[G,F] ,
\label{159}
\eeq
and acts like a derivation with the rules
\begin{eqnarray}
[F,GH] &=& [F,G]H + (-1)^{\epsilon_G(\epsilon_F+1)}G[F,H] \cr
[FG,H] &=& F[G,H] + (-1)^{\epsilon_G(\epsilon_H+1)}[F,H]G .
\end{eqnarray}

It is straightforward, but rather tedious, to check that it also satisfies
the super Jacobi identity
\beq
(-1)^{(\epsilon_F+1)(\epsilon_H+1)}[F,[G,H]]
+ {\mbox{\rm cyclic perm.}} = 0.
\label{161x}
\eeq
(The ingredients needed to show this are the super Lie equations (\ref{40}),
and
the Jacobi identity (\ref{45}) for the structure coefficients $c^i_{jk}$).

Although all of these relations are shared by the conventional antibracket,
the new bracket of course does not follow from the antibracket. On the
contrary, the bracket $[\cdot,\cdot]$ is more general than the antibracket,
to which it reduces in the trivial limit $u^A_i = \delta^A_i$:
\beq
[F,G] = (F,G) ~~~{\mbox{\rm when}} ~~~~u^A_i = \delta^A_i .
\eeq
Actually, the relation between the two brackets is slightly more general,
since the bracket $[\cdot,\cdot]$ can be reduced to the usual antibracket
$(\cdot,\cdot)$ through a redefinition of the antighosts $\phi^*_i$
whenever $u^A_i$ is invertible. This is, however, obviously a very special
case as well. The relations (\ref{159})-(\ref{161x}) define what has been
called a
Gerstenhaber algebra \cite{Penkava}.

Not surprisingly, it turns out that the classical Master Equation (\ref{115})
can be expressed in terms of the more general bracket:
\beq
[S^{BV},S^{BV}] = 0 .
\eeq

It is also interesting to consider the way the bracket acts on the fields
$\phi^A$ and antighosts $\phi^*_i$. Clearly, there can be no canonical
relations \`{a} la Poisson brackets since the number of fields $\phi^A$
in general is different from the number of antighosts $\phi^*_i$. Instead,
\beq
[\phi^A,\phi^*_i] = u^A_i ,
\eeq
while within this bracket the antighosts satisfy a (super)Lie algebra:
\beq
[\phi^*_i,\phi^*_j] = \phi^*_kc^k_{ij} .
\eeq
In addition, the bracket between two fields vanishes: $[\phi^A,\phi^B]=0$.

There are also relations between the $\bar{\Delta}$-operator and the new
bracket. Consider the way $\bar{\Delta}$ acts on a product:
\beq
\bar{\Delta}(FG) = F(\bar{\Delta}G) + (-1)^{\epsilon_G}(\bar{\Delta}F)G
+ (-1)^{\epsilon_G}[F,G] .
\eeq
As noticed by Witten \cite{Witten}, one can use this equation to {\em
define} the new bracket $[\cdot,\cdot]$, given $\bar{\Delta}$. This
approach is particularly useful in view of the fact that $\bar{\Delta}$
is nilpotent:
\beq
\bar{\Delta}^2 = 0 .
\eeq

A general theorem \cite{Penkava} then assures that both the super Jacobi
identity (\ref{161}) and the exchange relation are satisfied automatically.
Furthermore, using $\bar{\Delta}^2(FG) = 0$, one finds
\beq
\bar{\Delta}[F,G] = [F,\bar{\Delta}G] - (-1)^{\epsilon_G}[\bar{\Delta}F,G] .
\eeq
When the functional measure is invariant under $\phi^A \to g^A(\phi',a)$,
we can also
express the full quantum Master Equation entirely in terms of $\bar{\Delta}$
and the new bracket. From eq. (\ref{113}) it follows that in this case,
\beq
\frac{1}{2}[S^{BV},S^{BV}] = i\hbar\bar{\Delta}S^{BV} .
\label{169}
\eeq
It is interesting to note that, just as in the case of the conventional
antibracket, this equation can be compactly expressed as
\beq
\bar{\Delta}e^{\frac{i}{\hbar}S^{BV}} = 0 .
\eeq
A certain geometric interpretation \cite{Witten,H1} probably underlies
some of these observations. In the integration theory on the supermanifold
spanned by the fields $\phi^A$ and the antighosts $\phi^*_i$, the
operator $\bar{\Delta}$ should be viewed as a divergence operator
associated with
the invariant measure $[d\phi^*][d\phi]\rho(\phi)$.
In the language
of ref. \cite{Witten}, the quantum Master Equation (\ref{169}) implies that
$\exp[(i/\hbar)S^{BV}]$ can be considered as a ``closed form'', annihilated
by the exterior derivative defined by $\bar{\Delta}$. Similarly, the
BRST operator $\delta$, which in the formulation where the
ghosts $c^i$ have been integrated out, equals $\bar{\sigma}$ of eq.
(\ref{156}),
can conveniently be related to the $\bar{\Delta}$-operation.
Consider a theory without gauge symmetries, for which the connection
to the Ward identities is particularly simple. Once we let
$\bar{\sigma}$
act on an object $\phi^*_iF[\phi]$, we get (see also ref. \cite{HT}):
\beq
\bar{\sigma}[\phi^*_iF(\phi)] = \bar{\Delta}\left[\left(\frac{i}{\hbar}\right)
\phi^*_iF(\phi)e^{\frac{i}{\hbar}S^{BV}}\right] .
\eeq
This means that the Schwinger-Dyson BRST
Ward identities can be viewed as a generalized
Stokes theorem on the supermanifold. It is curious that there exists
an extension of the supermanifold (in which all antighost directions are
doubled by keeping the ghosts $c^i$) where this same statement is just a
reflection of an ordinary BRST Ward Identity. The reason for this
is to be found in the fact that the BRST
symmetry considered is precisely determined by the invariance properties of
the path integral measure. An equivalent expression for the divergence
operation in the formulation in which the ghosts $c^i$ are kept can presumably
be obtained by considering the transformation properties of all fields
$\phi^A, \phi^*_i, c^i$ under reparametrizations.

\section{Conclusions}

The aim of this paper has been to explore the more general framework in
which the Batalin-Vilkovisky Lagrangian BRST formalism is situated. Once
it is realized that this formalism can be derived from an underlying
principle, that of ensuring correct Schwinger-Dyson equations in the
path integral through the BRST symmetry \cite{us},
it becomes obvious how one can
generalize it in various different directions. For example, the original
Batalin-Vilkovisky construction \cite{Batalin} was concerned with
theories whose functional measures were invariant under arbitrary local
field shifts. The Schwinger-Dyson equations in those cases follow
precisely from exploiting this shift invariance. If instead the functional
measures are invariant under more general transformations -- e.g., motion
on a curved manifold -- one can derive from first principles the analogue
of the Batalin-Vilkovisky formalism for such cases by again promoting the
corresponding symmetry transformation to a BRST symmetry. By following
exactly the same procedure as in the flat case, we have shown how one
arrives at a new Lagrangian BRST scheme, this time covariant with respect
to transformations that leave the measure for the fields $\phi^A$ invariant.
The result coincides with the recent generalization of the
Batalin-Vilkovisky formalism that is inferred from general covariance
arguments on the supermanifold spanned by fields $\phi^A$ and antifields
$\phi^*_A$ alone \cite{Schwarz}. Here, instead, we are by construction
concerned only with covariance on the space of fields $\phi^A$, but this
is of course just a special case. We emphasize that this covariant
description of the Batalin-Vilkovisky formalism here is {\em derived} from
first principles. It is only by keeping also the ghosts $c^A$ that we can
uncover the underlying BRST principle behind this covariant generalization.
It reads
\begin{eqnarray}
\delta\phi^A &=& u^A_B(\phi)c^B \cr
\delta c^A &=& - \frac{1}{2}(-1)^{\epsilon_B}c^A_{BC}c^Cc^B \cr
\delta\phi^*_A &=& (-1)^{\epsilon_A}\frac{\delta^r S}{\delta\phi^B}
u^B_A(\phi) + i\hbar (-1)^{\epsilon_A+\epsilon_B}\bar{\Gamma}^B_{BA}
+ (-1)^{\epsilon_A\epsilon_B}\phi^*_M\bar{\Gamma}^M_{BA}c^B ,
\end{eqnarray}
in the non-Abelian formulation. As expected, this makes explicit reference
to both the action and the functional measure.

There is clearly no obstacle to considering also transformations that are
$\phi^*$-dependent, which would lead to the formulation that is fully
covariant on the supermanifold spanned by $\phi^A$ and $\phi^*_A$. But
we see no physical principle that could motivate such a mixing of fields
and antifields. From our point of view, the antifields $\phi^*_A$ are
just particular antighosts (remnants of the articially introduced
gauge symmetry associated with the given field redefinition),
and not more special or important than the
ghosts $c^A$. We might thus just as well consider transformations that
involve these ghosts $c^A$ too.

As we have shown, one can go further, and consider elevating an arbitrary
field transformation  (which does not necessarily leave the measure
invariant) into a BRST principle. The Ward identities of this BRST
symmetry will then be the set of relations among Green functions that
can be derived from the path integral by performing such a field
redefinition. In general, this will correspond to subsets of the full set
of Schwinger-Dyson equations. For this reason, it cannot in general be
used as a principle on which to base the quantization procedure.

Even when the path integral measure {\em is} invariant under the given
field transformation, this may not necessarily lead to the full set of
Schwinger-Dyson equations. This is because the object $u^A_i(\phi)$ that
enters into the corresponding BRST symmetry may not be invertible. When
this is the case, one arrives at a bracket structure $[\cdot,\cdot]$
which is distinct from the conventional antibracket $(\cdot,\cdot)$. We
have explored various properties of this new bracket in some detail, and
found that it shares a number of features with the usual
antibracket, to which it reduces when $u^A_i \!= \!\delta^A_i$.

The derivation of covariant Batalin-Vilkovisky quantization from
the underlying Schwinger-Dyson BRST symmetry can also be generalized
to the case of extended BRST symmetry \cite{BLT,DDJ}. It would be
most interesting to see how this how this compares with the recent
covariant formulation of Batalin, Marnelius and Semikhatov
\cite{BMS,ND}.

\vspace{0.5cm}

\noindent
{\sc Acknowledgement:}~ The work of J.A. has been partially supported by
Fondecyt 1950809,  El Programa del Gobierno Espa\~nol de
Cooperaci\'on Cient\'\i fica con Iberoam\'erica and
a collaboration CNRS-CONICYT.

\vspace{1.5cm}

\noindent
{\Large{\bf Appendices}}

\appendix
\section{Actions on Lie groups}

When discussing the case of theories with non-trivial measures, we implicitly
assumed that a natural choice of variables in the path integral would be
given independently of the (super) Lie group of transformations that left
the functional measure invariant. It often happens that the classical action
itself is expressed directly in term of group elements, and that we wish
to integrate over the left or right invariant measure on this group. It
is therefore of interest to see how the Schwinger-Dyson BRST symmetry,
the associated Master Equation for the extended action $S_{ext}$, and the
rest of the considerations above carry over to this case.

First, how do we determine the most general Schwinger-Dyson equations? One
way of answering this question is to find the appropriate generalization
of ``translations'' on a given (super) group manifold. We shall outline
a more general definition below, but let us first content ourselves with
the same approach as in section 2.2 above. That is, we shall explore the
identities that follow from using the fact that the functional measure
on the group is chosen to be either left or right invariant (or both, but
this is not needed). We shall for simplicity take the Lie group to be
free of Grassmann-odd directions.

The fields $\phi^A$ are thus taken to be matrices $U(x)$, elements of
the Lie group $G$. The notion of a Lie derivative $\nabla^a$ acting on group
elements is useful at this stage. One can choose it to be
\beq
\nabla^a \equiv i (Ut^a)_{ij}\frac{\delta}{\delta U_{ij}} ,
\eeq
where $t^a$ are the generators of the group, with $[t^a,t^b] = i
f^{ab}_c t^c$, and normalized to, say, $tr(t^at^b) =
\frac{1}{2}\delta^{ab}$. This Lie derivative acts much as an ordinary
derivative, with, $e.g.$, Taylor expansions of the form
\beq
f(Ue^{i\theta_at^a}) = f(U) + \theta_a\nabla^a f(U) + \frac{1}{2}
\theta_a \theta_b \nabla^a\nabla^b f(U) + \ldots
\eeq
Since Lie derivatives do not commute (but rather satisfy the Lie
algebra itself), care is required when more than one derivative is
involved.

Schwinger-Dyson equations for a theory of group elements $U(x)$,
action $S[U]$, and partition function
\beq
{\cal Z} = \int [dU] \exp\left[\frac{i}{\hbar}S[U]\right]
\eeq
take the form
\beq
\left\langle \nabla^a F[U] + \left(\frac{i}{\hbar}\right)
F[U]\nabla^a S[U]\right\rangle = 0 ,
\eeq
for an arbitrary function $F[U]$. These equations are Ward identities
of the equivalent theory based on \cite{us0}
\beq
{\cal Z} = \int [dU][d\phi^*][dc]\exp\left[\frac{i}{\hbar}(S[U] -
\phi^*_a c^a)\right] .
\label{157}
\eeq
We view what is in the exponent as the extended action, and denote it
by $S_{ext}$. The BRST symmetry under which both the action and the
right-invariant measure remains invariant reads
\begin{eqnarray}
\delta U(x) &=& iU(x)t_ac^a \cr
\delta c^a(x) &=& 0 \cr
\delta\phi^*_a &=& - \nabla_a S_{ext} ,
\label{158}
\end{eqnarray}
where in the last line we have used the decoupling of the ghost fields
from the classical action to rewrite the symmetry in terms of $S_{ext}$.
The Ward Identities of this symmetry are Schwinger-Dyson equations.

The above BRST symmetry suffers from not being nilpotent, even when just
acting on the group elements $U$. We can remedy
this by hand if we change the transformation law for the ghosts $c^a$. This
will not affect the Ward Identities $0\!=\!\langle\delta\{\phi^*_a F[U]
\rangle$, which hence still provide correct Schwinger-Dyson equations.
The required modification is
\beq
\delta c^a = \frac{1}{2}c^a_{bc}c^bc^c .
\eeq
Now $\delta^2 U \!=\!0$, as required, but the action in (\ref{157}) is
no longer
invariant under the BRST symmetry. We can again correct for this by hand
if we modify the transformation law for $\phi^*$ as well:
\beq
\delta\phi^*_a = -\nabla_aS_{ext} + \frac{1}{2}\phi^*_b c^b_{ca}c^c .
\eeq
On the surface this would seem to change the above Ward Identity. But when
we perform the ghost-antighost integrations, the extra term is seen to
contribute a term proportional to $c^a_{ab}$. This term hence vanishes
whenever the Lie group is semi-simple (a requirement we had to impose anyway,
because otherwise the integration measure for the ghosts $c^a$ would not
be invariant under the modified transformation law proposed above).

The above Schwinger-Dyson BRST symmetry implies that the extended
action satisfies a certain Master Equation. For the non-nilpotent
version (\ref{158}) it reads
\beq
\frac{\delta^r S_{ext}}{\delta\phi^*_a}\nabla_a S_{ext} = c^a\nabla_a
S_{ext} .
\label{161}
\eeq

It is clear that once a solution has been found, one can replace the
ghosts $c^a$ by $c^a + \alpha{\cal R}^a$, where $\alpha$ is an
arbitrary constant, and ${\cal R}^a$ is annihilated by $\nabla_a
S_{ext}$:
\beq
{\cal R}^a\nabla_a S_{ext} = 0 .
\eeq
Gauge symmetries involving solely fields that are elements of a
compact Lie group $G$ ordinarily need not be gauge fixed. But the
Schwinger-Dyson BRST symmetry (\ref{158}) is independent of
possible internal
gauge symmetries, and is valid in general --- as is the Master
Equation (\ref{161}) for such theories.
Since normally no other ghost fields will be present,
the solution (\ref{157}) suffices. However, one can conceive
of situations where gauge fixing is convenient, and where extra ghost
fields then have to be introduced, even in this context.\footnote{For
some recent examples, see, e.g., ref. \cite{Sollacher}.} These extra
fields will not be elements of the group $G$, and one therefore has to
specify additionally their BRST transformation law, on top of the
list given in eq. (\ref{158}). Once specified, one can immediately write down
the corresponding (quantum) Master Equation by demanding that the BRST
variation of the action is cancelled by that of the measure --- as was
done in section 3.

As this example has shown, the Master Equation can take quite
different forms depending on the field theory context. The antibracket
does not enter at all in the present case, and the Master Equation involves
fields $U(x) \in G$ that are on a quite different footing from their
natural partners in Batalin-Vilkovisky quantization, the antighosts
$\phi^*$ (which belong to the algebra of the group). Still, this is
the direct group theory analogue of the Batalin-Vilkovisky formalism.

\section{A more general Setting}

%Ref: Helgason, "Symmetric Spaces"
%E. Cartan: "Lessons sur la geometrie des espaces de Riemann"

Let ${\cal M}$ be a manifold and $G$ a group of transformations of
${\cal M}$. For convenience, let us here restrict ourselves to
bosonic manifolds.
We will say that $G$ acts transitively on
${\cal M}$ if given $x,y$ in ${\cal M}$ there exists an element
$g$ in $G$ such that
$$
y=g*x
$$
Actually, our future analysis will deal only with local properties of
${\cal M}$. So we will need a less restrictive property of $G$,
local transitivity. That
is, $ $G acts transitivily on every neighborhood of a given point of
${\cal M}$.

Let $x_0$ be a point of ${\cal M}$. The little group of
$x_0$ ($H_{x_0}$) is the subgroup
of elements in $G$ that do not change $x_0$.
We will refer to it intuitively as ``rotations".
i.e.:
$$
x_0=h*x_0 ~~~~,~~~   h\in H_{x_0}
$$

It is easy to show that if $G$ acts transitively in a certain
submanifold of ${\cal M}$, then $H$ is independent of the basis
point $x_0$ in that region. From now
on we will assume this to be the case, and will hence
not write explicitly the subindex of $H$.

Let ${\cal M}_0$ be a submanifold of ${\cal M}$
(the neighborhood of a point $x_0$ for instance)
and $G$ a group that acts transitively on ${\cal M}_0$.
Then it follows that ${\cal M}_0$ is isomorphic to $G/H$,
where $H$ is the little group of ${\cal M}_0$. We will refer to
the elements of $G/H$ as ``translations".
$G/H$ is called a symmetric space in the
literature. Since we only need the local property, we will refer to it as
locally symmetric space.

{}From now on we will study path integral quantization on locally symmetric
spaces. According to the last paragraph this covers a large class of
manifolds.

We will obtain Schwinger-Dyson equations only from ``translations". This
is the most general set of identities in cases of interest because normally
the action $S$ is invariant under ``rotations".
If this were not the case, then
the most general invariance of the measure
(both ``translations" and ``rotations") should be explored.
Then there will of course be more antifields than fields.

Our object of interest is the following functional integral:
\beqn
\int [dx]\rho(x) e^{-S[x]}
\eeqn

We will assume that the group of symmetries $G$ of the functional
measure acts transitively in a neighborhood of each point of
${\cal M}$. So we can choose as local coordinates the elements of
$G/H, ~x$. Moreover we can choose the integration
measure to be the Haar measure of $G$.

We apply now the collective coordinates method. Let $a$ be an element of
$G/H$. Then we consider:
\beqn
\int [da]\rho[a]\int [dx]\rho[x] e^{-S[a^{-1}*x]}
\eeqn

The new action is invariant under left multiplication,
\beqn
\tilde S[x,a] &=& S[a^{-1}*x]\cr
a'&=& b*a\cr
x'&=& b*x
\eeqn
In order to use the BRST method we need the infinitesimal transformation.
Let us choose variables such that $x=0$ corresponds to the identity of $G$.
$\phi(a,x)$ will be the (real) parameter corresponding to the element
$a*x$. This corresponds to the left multiplication rule.
Then the infinitesinal transformations are:
\beqn
\delta a_\alpha &=& \Theta_{\alpha\beta}(a)b^\beta \cr
\delta x_\alpha &=& \Theta_{\alpha\beta}(x)b^\beta \cr
\Theta_{\alpha\beta} &=& \left.\frac{\partial\phi(b,a)_\alpha}
{\partial b^\beta}\right|_{b=0}
\eeqn

The corresponding BRST transformation is:
\beqn
\delta a_\alpha &=& \Theta_{\alpha\beta}(a)c^\beta \cr
\delta x_\alpha &=& \Theta_{\alpha\beta}(x)c^\beta \cr
\delta c^\beta &=& \frac{1}{2} c^\beta_{\epsilon\gamma}
c^\epsilon c^\gamma \cr
\delta\bar c_\beta &=& i b_\beta \cr
\delta b_\beta &=& 0 ~.
\eeqn
Here
$c^\beta_{\epsilon\gamma}$ are the structure constants of the Lie group.
It is easy to check that the BRST generator correponding to
this set of transformations is nilpotent.

Now we fix for convenience the gauge $a=0$. The gauge fixed action is:
\beqn
\bar S=S[a^{-1}*x]-iba-\bar c_\alpha\Theta_{\alpha\beta}(a) c^\beta
\eeqn

Integration over $b$ and $a$ gives:
\beqn
S_{ext}=S(x)-\bar c_\alpha c^\alpha
\eeqn
which is invariant under

\beqn
\delta x_\alpha &=& \Theta_{\alpha\beta}(x)c^\beta \cr
\delta c^\beta &=& \frac{1}{2} c^\beta_{\epsilon\gamma}
c^\epsilon c^\gamma \cr
\delta\bar c_\beta &=& -\frac{\partial S}{\partial x^\sigma}
\Theta_{\sigma\beta}(x)+
\bar c_\alpha\left.\frac{\partial
\Theta_{\alpha\gamma}}{\partial a^\beta}\right|_{a=0}
c^\gamma ~.
\eeqn
The generalization to supermanifolds is straightforward.

\end{document}